\documentclass[journal,twoside]{IEEEtran}
\usepackage{graphicx}
\usepackage{color,soul}
\usepackage{multirow,booktabs}
\usepackage{cite}

\newcommand{\etal}{\textit{et al.}}
\usepackage{amsmath}
\usepackage{amsfonts}
\usepackage{amssymb}
\usepackage{cases}
\usepackage{bm}
\usepackage{algorithm}
\usepackage{algorithmic}
\usepackage{balance}
\usepackage{siunitx}
\usepackage{color}
\usepackage{bigstrut}
\usepackage{makecell}

\usepackage{array}

\ifCLASSOPTIONcompsoc
  \usepackage[caption=false,font=normalsize,labelfont=sf,textfont=sf]{subfig}
\else
  \usepackage[caption=false,font=footnotesize]{subfig}
\fi
 \let\MYorigsubfloat\subfloat
 \renewcommand{\subfloat}[2][\relax]{\MYorigsubfloat[]{#2}}
\usepackage{url}

\newcommand{\mathopr}[1]{\mathtt{#1}}

\usepackage{tikz,xcolor,hyperref}
\hypersetup{hidelinks}

\definecolor{lime}{HTML}{A6CE39}
\DeclareRobustCommand{\orcidicon}{%
	\begin{tikzpicture}
	\draw[lime, fill=lime] (0,0) 
	circle [radius=0.16] 
	node[white] {{\fontfamily{qag}\selectfont \tiny ID}};
	\draw[white, fill=white] (-0.0625,0.095) 
	circle [radius=0.007];
	\end{tikzpicture}
	\hspace{-3mm}
}

\foreach \x in {A, ..., Z}{%
	\expandafter\xdef\csname orcid\x\endcsname{\noexpand\href{https://orcid.org/\csname orcidauthor\x\endcsname}{\noexpand\orcidicon}}
}

\begin{document}
%
\title{Blind Super-Resolution for Remote Sensing Images via Conditional Stochastic Normalizing Flows}
%
%

\author{Hanlin~Wu\orcidA{},~Ning~Ni,~Shan~Wang,~Libao Zhang\orcidB{},~\IEEEmembership{Member,~IEEE}
    \thanks{
        This work was supported in part by the Beijing Natural Science Foundation under Grant L182029, in part by the National Natural Science Foundation of China under Grant 61571050 and Grant 41771407.
        \emph{(Corresponding author: Libao Zhang.)}

        The authors are with the School of Artificial Intelligence, Beijing Normal University, Beijing 100875, China. (e-mail: libaozhang@bnu.edu.cn).}
}

%
%

\markboth{}%
{}
%



\maketitle

\begin{abstract}
    Remote sensing images (RSIs) in real scenes may be disturbed by multiple factors such as optical blur, undersampling, and additional noise, resulting in complex and diverse degradation models. At present, the mainstream SR algorithms only consider a single and fixed degradation (such as bicubic interpolation) and cannot flexibly handle complex degradations in real scenes. Therefore, designing a super-resolution (SR) model that can cope with various degradations is gradually attracting the attention of researchers. Some studies first estimate the degradation kernels and then perform degradation-adaptive SR but face the problems of estimation error amplification and insufficient high-frequency details in the results. Although blind SR algorithms based on generative adversarial networks (GAN) have greatly improved visual quality, they still suffer from pseudo-texture, mode collapse, and poor training stability. In this article, we propose a novel blind SR framework based on the stochastic normalizing flow (BlindSRSNF) to address the above problems. BlindSRSNF learns the conditional probability distribution over the high-resolution image space given a low-resolution (LR) image by explicitly optimizing the variational bound on the likelihood. BlindSRSNF is easy to train and can generate photo-realistic SR results that outperform GAN-based models. Besides, we introduce a degradation representation strategy based on contrastive learning to avoid the error amplification problem caused by the explicit degradation estimation. Comprehensive experiments show that the proposed algorithm can obtain SR results with excellent visual perception quality on both simulated LR and real-world RSIs.
\end{abstract}


\ifCLASSOPTIONpeerreview
    \begin{center} \bfseries EDICS Category: 3-BBND \end{center}
\fi
%
\IEEEpeerreviewmaketitle

\section{Introduction}
%
%
%
%

\IEEEPARstart{R}{emote} sensing images (RSIs) are vulnerable to various factors such as sensor noise, imaging platform motion, and weather factors, resulting in the degradation of imaging quality. Super-resolution (SR) aims to restore clear texture details from low-resolution (LR) images, thereby improving the spatial resolution of RSIs. SR is a challenging ill-posed problem since different high-resolution (HR) images can be degraded into the same LR image through different degradation models.

\begin{figure}[t]
    \centering
    \includegraphics[width=\linewidth]{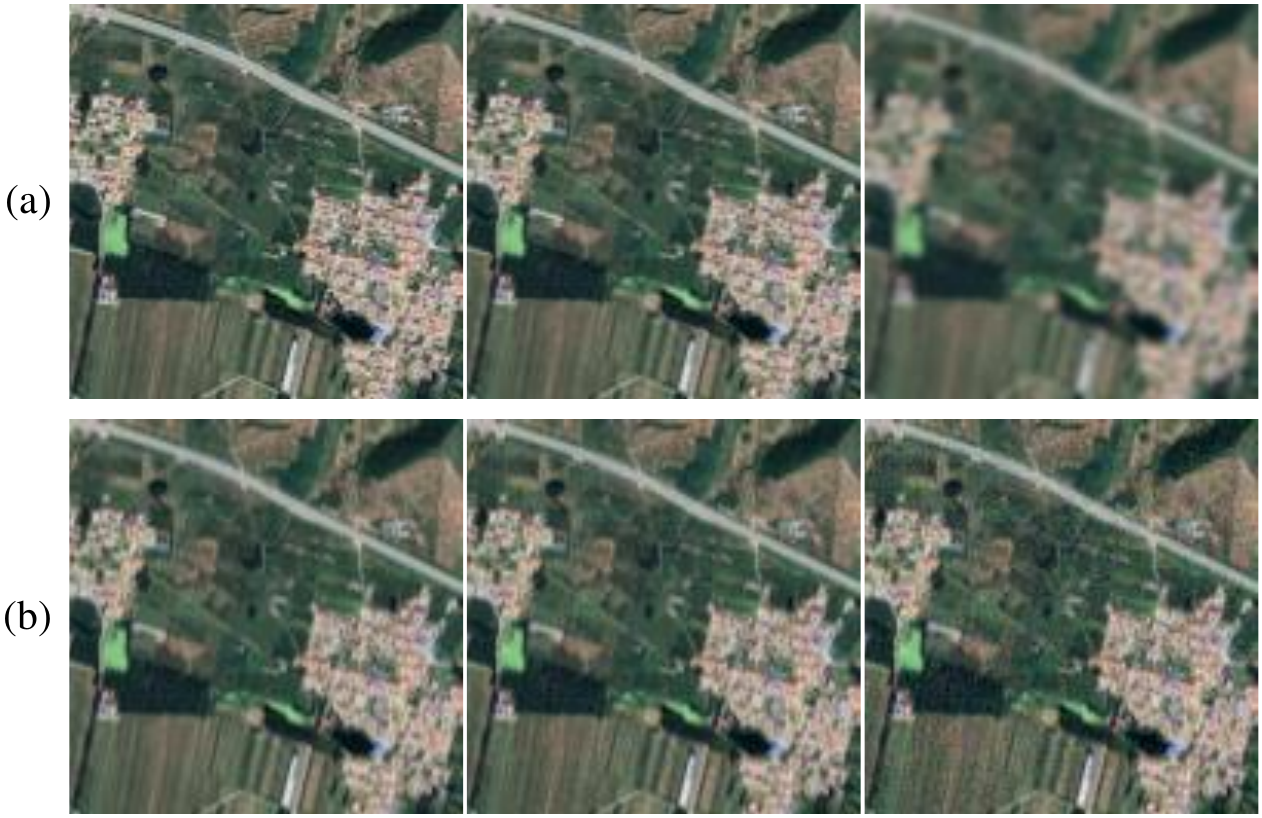}
    \caption{LR images generated by different degradations. (a) Various blur kernels. (b) Various noise levels.}
    \label{fig:intro}
\end{figure}

The convolutional neural network (CNN)-based SR algorithms have made great progress in objective metrics and visual perception quality \cite{yang2019deep, wang2020deep}. However, most methods assume that LR images are obtained by an ideal and fixed degradation model, such as bicubic interpolation, and thus cannot obtain satisfactory results in dealing with RSIs in real-world scenes. This is because the degradation models of RSIs are usually complex, and the LR images may suffer from various blur kernels and varying levels of noise, as shown in Fig.\,\ref{fig:intro}. When the degradation kernel used in the training phase mismatch the real LR images in the testing phase, the model will fail to generate satisfactory SR results. Generally, the degradation process of RSIs can be modeled as follows \cite{zhang2020multi}:
\begin{equation}
    \mathbf{X}_{\mathrm{LR}} = (\mathbf{X}_{\mathrm{HR}}\otimes k)\downarrow_r + n,
\end{equation}
where $\mathbf{X}_{\mathrm{LR}}$ and $\mathbf{X}_{\mathrm{HR}}$ denote the LR and HR image, respectively; the blur kernel $k$ and the additive noise $n$ are two key factors of the degradation process; $\downarrow_r$ denotes a downsampling operation with a scale factor of $r$.
In this case, it is impractical to train a model for each degradation kernel, which will cost huge model training and storage resources.
Therefore, it is necessary to build a degradation-adaptive SR algorithm for RSIs in real-world scenes.

Recently, some studies have focused on handling multiple degradations using a single model in real-world SR tasks, which can be categorized into \emph{blind SR} and \emph{non-blind SR}.
The non-blind SR models \cite{zhang2018learning,zhang2020deep,xu2020unified} rely on the real degradation information and the LR image together as input in the testing phase.
However, real degradation models in practical applications are often complex, unknown and difficult to obtain.
To fill this gap, blind SR models do not require degradations as priors.
Previous blind SR methods \cite{huang2020unfolding,he2021srdrl} usually decompose the task into two consecutive steps, degradation kernel estimation \cite{michaeli2013nonparametric,bell2019blind} and non-blind SR.
However, the SR step may amplify the estimation error of degraded kernels, leading to poor SR results.
Gu \etal \cite{gu2019blind} propose to iteratively estimate degenerate kernels and perform SR, which alleviates the difficulty of degradation kernel estimation at the cost of high computational complexity.

The aforementioned blind SR methods are optimized using pixel-level losses. Although these methods can obtain satisfactory PSNR, they tend to generate blurred SR results with poor visual quality.
These pixel-level loss-optimized methods learn a deterministic mapping from LR images to HR images, ignoring the ill-posed nature of SR tasks \cite{lugmayr2020srflow}.
Recent studies \cite{ji2020realworld,zhang2021designing,wang2021real} have introduced adversarial learning into the blind SR task.
They learn the probability distribution of HR space and obtain much clearer SR results.
Nonetheless, these methods based on generative adversarial networks (GANs) still deterministically map LR images to SR results, and do not inherently alleviate the ill-posed problem.
Furthermore, due to the convergence difficulties of GAN-based models, the generated results may suffer from model collapse.

Normalizing flow \cite{rezende2015variational} is another important class of generative models besides GAN.
Lugmayr \etal \cite{lugmayr2020srflow} proposed to use a flow-based generative model to explicitly learn the probability distribution over the HR image space, from which multiple SR results can be sampled.
However, the flow-based model requires the network to be invertible, which greatly increases the difficulty of architecture designing and limits the expressiveness of the network.
Lately, researchers \cite{ho2020denoising,song2020denoising,wu2020stochastic} employed Gaussian diffusion process to generalize the normalizing flow to the stochastic case, named stochastic normalizing flow (SNF). SNF does not require the network to be invertible and has demonstrated superior performance in many applications such as 3D cloud point generation \cite{luo2021diffusion} and speech synthesis \cite{jeong2021diff}.

In this article, we propose a novel blind SR model based on SNF (BlindSRSNF) to address the severe ill-posedness of blind SR tasks, the lack of texture details in SR results, and the instability of GAN-based model training. BlindSRSNF utilizes a Markov process to transform the distribution from the HR image space to a Gaussian latent space. Then, we take the LR image encoding and degradation information as conditions to construct the conditional transition probability of the reverse Markov process. This reverse process maps the samples in the latent space to the HR image space by transferring hidden variables step by step, which can be regarded as the sample generation process. Furthermore, to address the problem of inaccurate estimation of degradation kernels, we introduce a degradation representation strategy \cite{wang2021unsupervised} based on unsupervised contrastive learning. Therefore, the proposed model is robust to various degradation models and can achieve satisfactory blind SR results in real scenarios.

To the best of our knowledge, BlindSRSNF is the first SNF-based blind SR method that provides a new idea for improving the quality of RSIs in real scenes. The main contributions of this article are as follows:

\begin{enumerate}
    \item We propose a novel SNF-based blind SR framework for RSIs named BlindSRSNF, which can effectively stabilize model training by explicitly optimizing the variational bound of the NLL.
    \item The BlindSRSNF adopts contrastive learning to learn the degradation information of LR images in an unsupervised manner, avoiding the amplification of the degradation kernel estimation error and the time-consuming iterative degradation correction.
    \item Compared with the state-of-the-art (SOTA) GAN-based blind SR algorithms, our proposed BlindSRSNF can generate SR results with better visual perception quality. Our results have more natural, accurate texture details with lower spatial distortion.
\end{enumerate}

\section{Related Works}
\label{sec:related_works}

\subsection{Bicubic-Assumed SR}

Recently, with the rapid development of deep learning (DL), SISR have made great progress. Many DL-based methods were proposed to learn the mapping from LR space to HR space in an end-to-end manner. Dong \etal \cite{dong2015image} proposed the first DL-based end-to-end SR network, which greatly improved the performance of SISR. Kim \etal \cite{kim2016accurate} proposed a residual network, which can efficiently learn the high-frequency information of images and reduce learning burden of the network. Ledig \etal \cite{ledig2017photo} proposed a generative adversarial network, it can generate more realistic details and textures, which greatly improved visual perceptual quality of SR. Zhang \etal \cite{Zhang2018d} proposed a residual dense network (RDN), which uses the dense and skip structure and further improves the SR performance. Zhang \etal proposed a deep residual channel attention network (RCAN), which consists of several residual groups with long skip connections. The RCAN embeds a channel attention mechanism, which can  adaptively rescale channel-wise features by considering interdependencies among channels. Li \etal \cite{Li2019c} introduced a feedback mechanism into the SR task and proposed a lightweight super-resolution feedback network. Dai \etal \cite{dai2019second} proposed a second-order attention network for more powerful feature expression and feature correlation learning. Soh \etal \cite{soh2020meta} proposed a novel training scheme based on meta-transfer learning to exploit both external information from a large-scale dataset and internal information from a specific image. Kong \etal \cite{kong2021classsr} found that different image regions have different restoration difficulties and proposed a new solution pipeline, which makes different regions be processed by networks with different capacities and reduces a lot of computational consumption.

Ma \etal \cite{ma2019achieving} employed a wavelet transform and recursive res-net to achieve single image super-resolution for remote sensing images. Arun \etal \cite{arun2020cnn} explored an optimal spectral super-resolution framework for remote sensing images, which can ensure the spectral and spatial fidelity of reconstructions with mini-mum number of samples. Lei \etal \cite{lei2020coupled} presented a coupled adversarial training mode for remote sensing image super-resolution, in which the discriminator is specifically designed to take in a pair of images rather than a single input to make better discrimination of the inputs. Zhang \etal \cite{zhang2020scence} proposed a multiscale attention network to characterize the structural features of remote sensing images at multiple levels for remote sensing image super-resolution. Huan \etal \cite{huan2021end} proposed an improved multi-scale residual network, which combined hierarchical residual-like connections and dilation convolution to solve the problem of forgetting and underutilizing network features. Wu \etal \cite{wu2022remote} used the saliency-guided feedback GAN to discriminate different regions with varying levels of saliency and reconstruct the high-resolution remote sensing images. Lei \etal \cite{lei2022hybird} proposed a hybrid-scale self-similarity exploitation network (HSENet) to learn single- and cross-scale internal recurrence of patterns in remote sensing images. Li \etal \cite{li2022signle} designed an attention-based GAN model that applied both local attention and global attention for the super-resolution task of remote sensing images.

\subsection{Degradation-Adaptive SR}

Shocher \etal \cite{shocher2018zero} trained a small CNN suitable for the test images to adapt to the specific degraded kernel in the test stage, but the inference efficiency of the model is low. Zhang \etal \cite{zhang2018learning} took the fuzzy kernel information and LR images as the input of the network, and proposed a SR network for multiple degradations (SRMD). After that, Zhang \etal \cite{zhang2020deep} proposed an end-to-end unfolding SR network (USRNet) to deal with different degradations by alternately solving data and prior problems. Xu \etal \cite{xu2020unified}, based on dynamic convolution, further improved the performance of SR under a variety of degraded kernels. However, these non-blind SR reconstruction algorithms \cite{zhang2018learning,zhang2020deep,xu2020unified} rely on the degraded information provided by the degraded kernel estimation methods \cite{bell2019blind,michaeli2013nonparametric} to perform SR tasks. In addition, the SR network will magnify the estimation errors of degraded kernel, causing obvious artifacts in the results \cite{gu2019blind}.

To solve this problem, Gu \etal \cite{gu2019blind} proposed an iterative kernel correction (IKC) method, which used the SR reconstruction results of the previous iteration to correct the degraded kernel estimation, so as to improve the SR quality of the next iteration. Although IKC can effectively alleviate the artifact problem caused by degraded kernel estimation error, multiple iterations in the test stage are very time-consuming.

\subsection{Flow-Based Models}

Normalizing Flow (NF)\cite{dinh2014nice,rezende2015variational,papamakarios2021normalizing} is a kind of generative model, which has a wide range of applications in nuclear physics, materials science and other fields\cite{wu2020stochastic,albergo2019flow,li2018neural}. In recent years, in the field of computer vision, it is gradually attracting the attention of researchers\cite{lugmayr2020srflow, kingma2018glow, dinh2016density}.

To solve above problem, some researchers introduce normalizing flow (NF) \cite{rezende2015variational} into SR reconstruction tasks. Lugmayr \etal \cite{lugmayr2020srflow} proposed a SR with normalizing flow (SRFlow), which used NF to model the conditional distribution in HR space. In this way, SRFlow can directly optimize the negative log-likelihood function to obtain realistic SR reconstruction results. Compared with the GAN-based methods, the NF-based models can effectively avoid the problems of mode collapse \cite{isola2017image,mathieu2015deep} and unstable training \cite{arjovsky2017wasserstein}. However, the NF model requires each layer of the neural network to be reversible so as to establish bijection from HR space to hidden space. The conventional convolution layer is difficult to meet the requirements of reversibility, and the specially designed reversibility layer will greatly limit the expression ability of the network \cite{wu2020stochastic}. Subsequently, Ho \etal \cite{ho2020denoising} proposed a denoising diffusion probabilistic model, which mapped the samples to the hidden space using the diffusion process, and constructed an inverse denoising process to generate samples. This kind of generation models \cite{ho2020denoising,song2020denoising} that realizes probability distribution transformation based on stochastic process is also called stochastic normalizing flow (SNF).

\section{Preliminary}
\label{sec:preliminary}

\subsection{Normalizing Flow}
The NF is a series of invertible functions parameterized by a neural network to realize the probability distribution transformation from a prior space $\mathcal Z$ to the target space $\mathcal X$. We can obtain the exact probability distribution over the target space by using the change of variable theorem. Thus, the parameters of the neural network can be optimized by maximizing the likelihood of samples or minimizing the Kullback-Leibler (KL) divergence of the generated distribution from the target distribution.

Let $F_{ZX}$ denote the invertible mapping from the space $\mathcal Z$ to $\mathcal X$. To simplify the construction of the invertible function, $F_{ZX}$ is usually decomposed into $T$ invertible layers $F_0,\cdots, F_T$,
\begin{equation}
    \mathbf{y}_{t+1} = F_t(\mathbf{y}_t),\ \ \mathbf{y}_t = F_{t}^{-1}(\mathbf{y}_{t+1}),
\end{equation}
where $\mathbf{y}_t, 0\leq t\leq T$ are intermediate states. Let $\mathbf z$ and $\mathbf{x}$ denote samples in $\mathcal Z$ and $\mathcal X$ spaces, respectively, then the NF model can be expressed as:
\begin{equation}
    \mathbf{z}=\mathbf{y}_{0} \underset{F_0^{-1}}{\stackrel{F_{0}}{\rightleftarrows}} \mathbf{y}_{1} \rightleftarrows \cdots \rightleftarrows \mathbf{y}_{T-1} \underset{F_{T-1}^{-1}}{\stackrel{F_{T-1}}{\rightleftarrows}} \mathbf{y}_{T}=\mathbf{x}.
\end{equation}
Assuming that each transformation function is differentiable, let $|\operatorname{det} \mathbf{J}_{t}(\mathbf y)|$ denote its Jacobian determinant. Using the change of variable theorem, we can calculate the probability density of $\mathbf {y}_{t+1}$,
\begin{equation}
    p_{t+1}\left(\mathbf{y}_{t+1}\right)=p_{t+1}\left(F_{t}\left(\mathbf{y}_{t}\right)\right)=p_{t}\left(\mathbf{y}_{t}\right)\left|\operatorname{det} \mathbf{J}_{t}\left(\mathbf{y}_{t}\right)\right|^{-1}.
\end{equation}
Then, the negative log-likelihood (NLL) of the training samples can be obtained,
\begin{equation}
    \mathcal L(\Theta;\mathbf{x}) = -\log p_{\mathbf{x}}(\mathbf{x}) = -\log  p_{\mathbf{z}}(\mathbf{z}) - \sum_{t=0}^{T-1} |\operatorname{det} \mathbf{J}_{t} (\mathbf{y}_{t})|.
\end{equation}
Thus, the network parameters can be optimized by minimizing the NLL by computing the Jacobian determinant of each transformation function.

In practice, each transformation function corresponds to a layer in the neural network. For efficient training and inference, the inverse and Jacobian determinants of each layer must be efficiently computed. However, designing a layer that satisfies the above characteristics is challenging because common neural network structures are not invertible. The literature \cite{dinh2014nice,dinh2016density} proposes an affine coupling layer, which provides a simple and effective way for constructing an invertible neural network layer. However, due to the limitation of invertibility, the expressive ability of the network is severely constrained \cite{wu2020stochastic}.

\subsection{Stochastic Normalizing Flow}

\begin{figure*}[t]
    \centering
    \includegraphics[width=\textwidth]{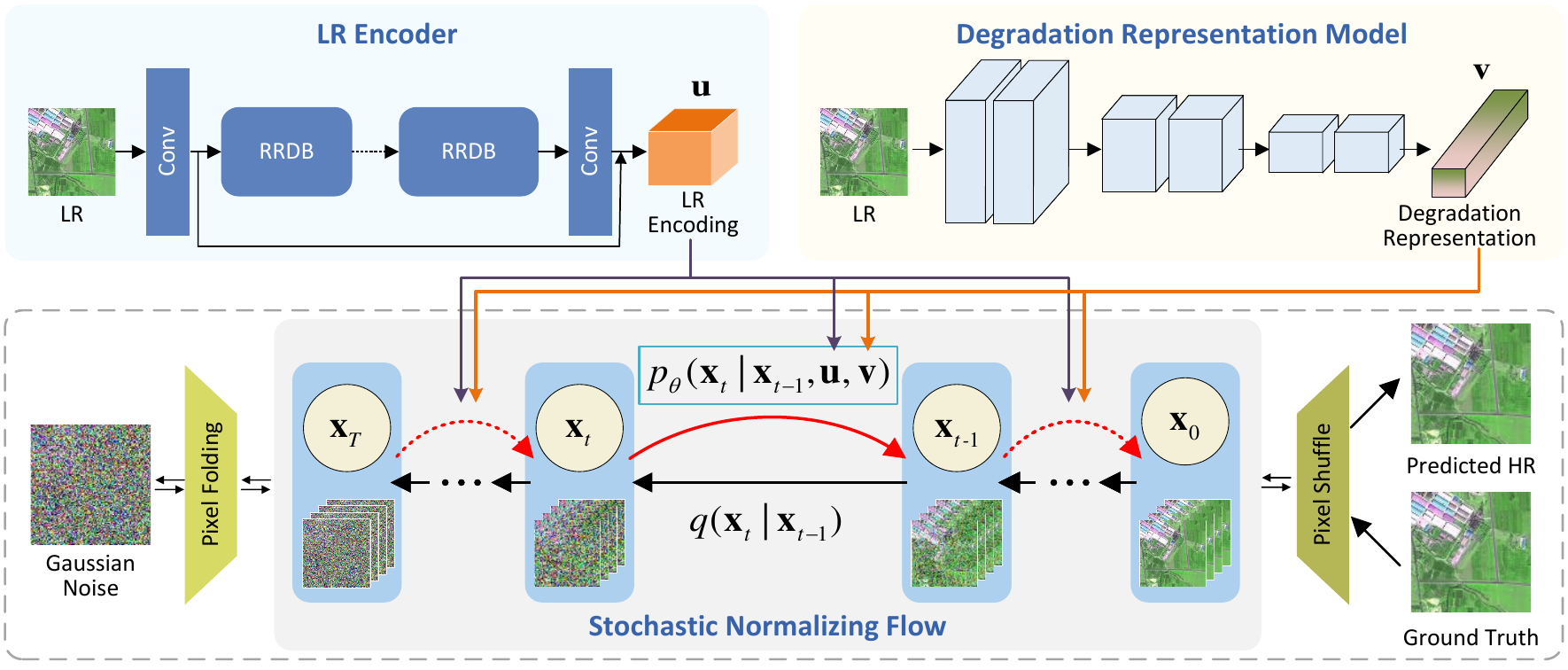}
    \caption{Flowchart of the proposed BlindSRSNF.}
    \label{fig:flowchart}
\end{figure*}

The SNF is a generalization of the NF in the random case, which realizes the transformation of the hidden state through random sampling, rather than a fixed invertible function. Unlike the NF, which implements probability distribution transformation through $T$ certain functions, the SNF constructs a Markov chain $\{\mathbf X_t\}_{t=0}^T$ of length $T+1$ to achieve probability distribution transformation. From the perspective of stochastic processes, the NF can also be regarded as a special case where all transformation probabilities in the Markov chain are \emph{Dirac measure} (probability mass is concentrated at a single point).

Let $\mathbf X_T$ denote standard Gaussian noise. The SNF starts from $\mathbf X_{T}$ and passes through $T$-step probability transformation to obtain the sample $\mathbf X_0$ in the target space. This process of sample generation is also called \emph{reverse process}. Conversely, starting from $\mathbf X_0$, the process of corrupting the sample to get random Gaussian noise is called \emph{forward process}. To simplify the model, we assume the forward process is a \emph{diffusion process} \cite{ho2020denoising}, that is, Gaussian noise is gradually added to the sample according to the variance sequence $\beta_1\cdots, \beta_T$. Specifically, let $\bm I$ denote the identity matrix, and the joint probability and transition probability of the forward process are respectively defined as:
\begin{align}
    q(\mathbf X_{1:T}|\mathbf X_0) & := \prod_{t=1}^{T}q(\mathbf X_{t}|\mathbf X_{t-1}),                      \\
    q(\mathbf X_t|\mathbf X_{t-1}) & :=\mathcal{N}(\mathbf X_t;\sqrt{1-\beta_t}\mathbf X_{t-1},\beta_t\bm I).
\end{align}
It is worth noting that, let $\alpha_t := 1-\beta_t$, $\bar \alpha_t := \prod_{s=1}^t\alpha_s$, the $t$-step transition probability of the forward process can be calculated exactly:
\begin{equation}
    \label{eq:aZs2Rh}
    q\left(\mathbf{X}_{t} \mid \mathbf{X}_{0}\right)=\mathcal{N}\left(\mathbf{X}_{t} ; \sqrt{\bar{\alpha}_{t}} \mathbf{X}_{0},\left(1-\bar{\alpha}_{t}\right) \bm I\right).
\end{equation}
The variances $\{\beta_t\}_{t=1}^T$ of the forward process are hyperparameters. When the variances $\{\beta_t\}_{t=1}^T$ are small, the transition probability of the reverse process $p_\theta(\mathbf{X}_{t-1} | \mathbf{X}_t)$ also obeys Gaussian distribution \cite{sohl2015deep}, where $\theta$ denotes the model parameters. Therefore, the reverse process is a Markov chain whose transition probability follows a Gaussian distribution, and the initial distribution $p(\mathbf{X}_T) = \mathcal N(\mathbf{X}_T; \bm 0, \bm I)$. Specifically,
\begin{align}
    p_\theta(\mathbf X_{T-1:0} |\mathbf{X}_T)                   & := \prod_{t=1}^{T}p_\theta(\mathbf X_{t-1}|\mathbf X_t),
    \label{eq:fKxa}                                                                                                                                                                                                 \\
    p_{\theta}\left(\mathbf{X}_{t-1} \mid \mathbf{X}_{t}\right) & :=\mathcal{N}\left(\mathbf{X}_{t-1} ; \bm{\mu}_{\theta}\left(\mathbf{X}_{t}, t\right), \bm{\Sigma}_{\theta}\left(\mathbf{X}_{t}, t\right)\right).
\end{align}
The training of the generative model $p_\theta(\mathbf X_0)$ can be achieved by optimizing the variational upper bound of the NLL,
\begin{equation}
    \label{eq:jONtBG}
    \begin{aligned}
             & \mathbf E[-\log p_\theta(\mathbf{X}_0)]                                                                                                          \\
        \leq & \mathbf E_q\left[-\log \frac{p_\theta(\mathbf{X}_{T:0})}{q(\mathbf{X}_{1:T}|\mathbf{X}_0)}\right]                                                \\
        =    & \mathbf E_q\left[-\log p(\mathbf{X}_T)-\sum_{t\geq 1}\log\frac{p_\theta(\mathbf{X}_{t-1}|\mathbf{X}_t)}{q(\mathbf{X}_t|\mathbf{X}_{t-1})}\right]
        = :  L.
    \end{aligned}
\end{equation}

We adopt the parameterization method proposed by Ho \etal \cite{ho2020denoising}, where $\Sigma_\theta(\mathbf{X}_t, t) = \sigma_t^2\bm I$, and the mean term has the following form:
\begin{equation}
    \label{eq:NfjtNt}
    \boldsymbol{\mu}_{\theta}\left(\mathbf{X}_{t}, t\right)=\frac{1}{\sqrt{\alpha_{t}}}\left(\mathbf{X}_{t}-\frac{\beta_{t}}{\sqrt{1-\bar{\alpha}_{t}}} \boldsymbol{\epsilon}_{\theta}\left(\mathbf{X}_{t}, t\right)\right),
\end{equation}
where $\bm\epsilon_{\theta}$ is a function to predict noise $\bm \epsilon$, accepting $\mathbf{X}_t$ and $t$ as input. The optimization objective $L$ in \eqref{eq:jONtBG} can be simplified to the following form:
\begin{equation}
    \label{eq:5yAs5W}
    \min _{\theta} \mathcal L_{\mathrm{simple}}(\theta)=\mathbf{E}_{\mathbf{X}_{0}, \bm\epsilon, t}\left\|\bm\epsilon-\bm\epsilon_{\theta}\left(\sqrt{\bar{\alpha}_{t}} \mathbf{X}_{0}+\sqrt{1-\bar{\alpha}_{t}} \bm\epsilon, t\right)\right\|^{2},
\end{equation}
where $\bm\epsilon\sim \mathcal N(\bm 0, \bm I)$ is random noise.

\begin{figure*}[t]
    \includegraphics[width=\textwidth]{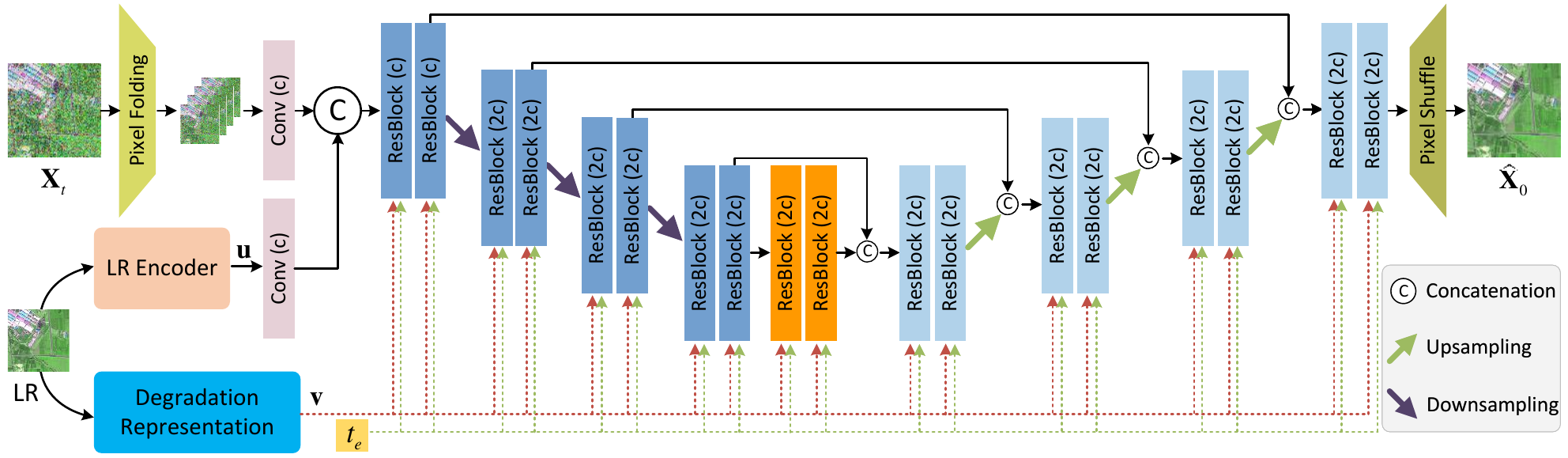}
    \caption{Architecture of the Denoising Network}
    \label{fig:unet}
\end{figure*}

\section{Proposed Method}
\label{sec:methodology}

In this section, we first give an overview of our proposed conditional SNF for blind SR (Sec.\,\ref{sec:conditional_snf}), and then introduce the specific parameterization method (Sec.\,\ref{sec:denoising_model}). Then, the LR encoder and degradation representation model are detailed in Sec.\,\ref{sec:LR_encoder} and Sec.\,\ref{sec:degradation_representation}. Finally, we summarize the training and inference process of the proposed model in Sec.\,\ref{sec:training} and Sec.\,\ref{sec:inference}.

\subsection{Conditional SNF for Blind SR}
\label{sec:conditional_snf}

The goal of the SR task is to generate an HR image $\mathbf{X}_{\mathrm{HR}}$ given an LR image $\mathbf{X}_{\mathrm{LR}}$. Since the mapping from LR images to HR images is one-to-many, we propose to parameterize the conditional distribution $p(\mathbf{X}_{\mathrm{HR}} | \mathbf{X}_{ \mathrm{LR}})$ based on the SNF model. In order to adapt the SR model to multiple degradation kernels, the generative model should comprehensively consider the content information of LR images and degradation representation. Therefore, we construct a conditional SNF model, in which the transition probability of the reverse process is modeled as the conditional probability given the LR encoding and degradation representation vector. The reverse process starts from Gaussian noise, and after $T$-step probability transformation, a sample $\mathbf{X}_0$ that obeys the conditional distribution $p(\mathbf{X}_{\mathrm{HR} } | \mathbf{X}_{\mathrm{LR}})$ can be obtained, thereby realizing blind SR of RSIs. The overall framework is shown in Fig.\,\ref{fig:flowchart}.

The LR encoder $f_\theta$ aims to extract the content information of the LR images as the condition of the SNF to ensure the consistency of the SR result with the LR image. The degradation representation model $g_\theta$ aims to extract a vector $\mathbf{v}$ from LR images that can effectively characterize its degradation information. Let $\theta$ be the set of all learnable parameters of the proposed BlindSRSNF. We take the LR encoding $\mathbf u = f_\theta(\mathbf{X}_{\mathrm{LR}})$ and the degradation representation vector $\mathbf v = g_\theta(\mathbf{X}_{\mathrm{LR}})$ as conditions and define the the transition probability of the conditional SNF. The reverse process is defined as:
\begin{equation}
    \begin{aligned}
        p_\theta(\mathbf X_{T-1:0}|\mathbf{X}_T)                                            & := \prod_{t=1}^{T}p_\theta(\mathbf X_{t-1}|\mathbf X_t, \mathbf u, \mathbf v),                                                                \\
        p_{\theta}\left(\mathbf{X}_{t-1} \mid \mathbf{X}_{t}, \mathbf{u}, \mathbf{v}\right) & :=\mathcal{N}\left(\mathbf{X}_{t-1} ; \bm{\mu}_{\theta}\left(\mathbf{X}_{t}, t, \mathbf{u}, \mathbf{v} \right), \bm{\sigma}_t^2 \bm I\right).
    \end{aligned}
\end{equation}
The initial distribution of the reverse process $p(\mathbf{X}_T) := \mathcal N(\mathbf{X}_T; \bm 0, \bm I)$.

It can be seen from \eqref{eq:NfjtNt} that the reverse process is determined by the noise prediction function $\bm \epsilon_\theta$. Therefore, we only need to model $\bm \epsilon_\theta$, and we obtain a generative model. According to \eqref{eq:aZs2Rh}, we have
\begin{equation}
    \mathbf{X}_t=\sqrt{\bar\alpha_t}\mathbf X_{\mathrm{HR}}+\sqrt{1-\bar\alpha_t}\bm\epsilon,\bm\epsilon\sim \mathcal{ N}(\bm0, \bm I), \ 1<t\leq T.
\end{equation}
If a model can accurately predict the noise $\bm\epsilon$ according to $\mathbf{X}_t$, it is equivalent to predicting the ``clean'' image before adding noise $\mathbf{X}_{\mathrm{HR}}$. Therefore, in our blindSRSNF, we construct a denoising model $h_{\theta}$ to parameterize the inverse process of the SNF. The goal of the denoising model is to remove the noise added by the forward diffusion process based on the LR encoding and degradation representation vector and obtain a ``clean'' predicted image $\mathbf{X}_0$. The inputs of the denoising model include $\mathbf{X}_t$, the time step of the diffusion process $t$, the feature encoding of the LR image $\mathbf u$ and the degradation representation vector $\mathbf v$, and the output is denoted as $\mathbf{\hat X}_0$.

In this study, we use $L_1$ norm instead of the $L_2$ norm as the loss function of the denoising model for better convergence performance. Therefore, the loss function of the stochastic normalized flow model is defined as:
\begin{equation}
    \label{eq:hWJnsO}
    \begin{aligned}
        \mathcal L_{\mathrm{SNF}}  = & \mathbf E_{\mathbf{X}_{\mathrm{LR}}, \mathbf{X}_{\mathrm{HR}}} \mathbf E_{\bm\epsilon, t}\|\mathbf{X}_{\mathrm{HR}} - \\ & h_\theta(\sqrt{\bar\alpha_t}\mathbf{X}_{\mathrm{HR}}+\sqrt{1 -\bar\alpha_t}\bm\epsilon, t, \underbrace{f_\theta(\mathbf{X}_{\mathrm{LR}})}_{\mathbf{u}}, \underbrace{g_\theta (\mathbf{X}_{\mathrm{LR}}}_{\mathbf{v}})\|_1.
    \end{aligned}
\end{equation}

\subsection{Architecture of the Denoising Network}
\label{sec:denoising_model}

\begin{figure}[t]
    \centering
    \includegraphics[width=0.8\linewidth]{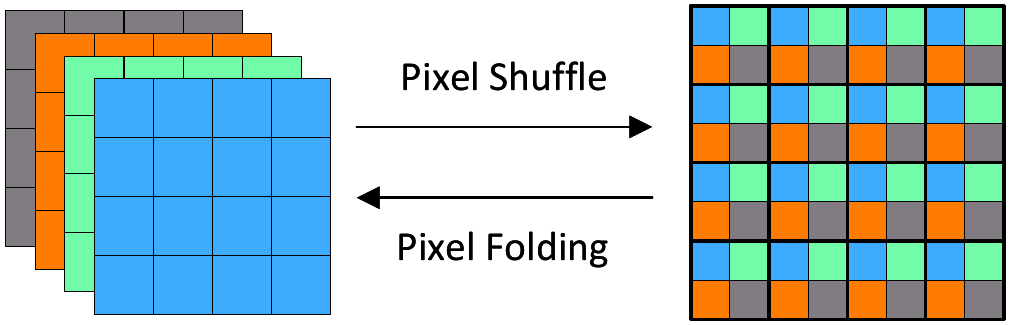}
    \caption{Illustration of the pixel folding and pixel shuffle operations.}
    \label{fig:pixelshuffle}
\end{figure}

We take U-Net \cite{ronneberger2015u} as the backbone of the denoising model $h_{\theta}$, as shown in Fig.\,\ref{fig:unet}.
To reduce the computational complexity of the reverse process, we use the pixel folding operation to reduce the spatial resolution of images.
The pixel folding is the inverse operation of pixel shuffle, as shown in Fig.\,\ref{fig:pixelshuffle}.
The pixel folding operation rearranges the pixels and reduces the spatial resolution to $1/2$ of the original image without losing information.

We take a convolutional layer to extract shallow features from the pixel-folded image, which are then concatenated with the LR feature encoding $\mathbf{u}$ and fed into the U-shaped network.
The compression path of the U-Net contains four feature extraction groups, each of which consists of two residual blocks and a downsampling operation.
The downsampling operation is implemented by a $3\times 3$ convolutional layer with a stride of $2$.
The expansion path of the U-Net contains four feature extraction groups, each preceded by an upsampling layer to increase the resolution of the feature maps.
The upsampling operation is implemented by the nearest neighbor interpolation and a $3\times 3$ convolutional layer.
Then, we concatenate the feature maps of the same size in the expansion path and the compression path and feed them into two residual blocks.
Each residual block accepts not only feature maps as input, but also a time encoding $t_e$ of step $t$ and a degradation representation vector $\mathbf v$ as conditions.
The time encoding $t_e$ is defined as:
\begin{equation}
    \begin{aligned}
        \phi(t) & = [\sin(\omega_1 t), \cos(\omega_1 t), \sin(\omega_2 t), \cos(\omega_2 t),\cdots ], \\
        t_e     & = \mathopr{MLP_3}(\phi(t)),
    \end{aligned}
\end{equation}
where $\{\omega_1,\omega_2,\cdots\}$ are frequency parameters, $\mathopr{MLP_3}(\cdot)$ represents a three-layer MLP where the hidden layer uses Swish \cite{ramachandran2017searching} as the activation function.

The residual block contains two convolution layers, each preceded by a group normalization \cite{wu2018group} to stabilize the training, using Swish as the activation function. To enable the residual block to perceive the degradation information of the LR image and dynamically adjust the kernel weights, we design a degradation-aware convolution (DAConv) layer to replace the second ordinary convolutional layer in the residual block. The DAConv layer takes the degradation representation vector as an additional input, which first uses a three-layer MLP to calculate the kernel weights, and then uses the kernel to perform the convolution operation. Let $\mathopr{DAConv}(\cdot, \mathbf{v})$ denote the DAConv layer; $\mathopr{GroupNorm}(\cdot)$ denotes the group normalization; $F_{\mathrm{in}}$ and $F_{\mathrm{out}}$ denote the input and output of the residual block, respectively. Then, the operation of the residual block can be summarized as:
\begin{equation}
    \begin{aligned}
        F_1              & = \mathopr{Conv}_{3\times 3}(\mathopr{Swish}(\mathopr{GroupNorm}(F_{\mathrm{in}}))),                       \\
        F_2              & = F_1 + t_e,                                                                                               \\
        F_{\mathrm{out}} & = \mathopr{DAConv}_{3\times 3}(\mathopr{Swish}(\mathopr{GroupNorm}(F_{2})), \mathbf{v}) + F_{\mathrm{in}}.
    \end{aligned}
\end{equation}

\subsection{LR Encoder}
\label{sec:LR_encoder}

\begin{figure}[t]
    \centering
    \includegraphics[width=\linewidth]{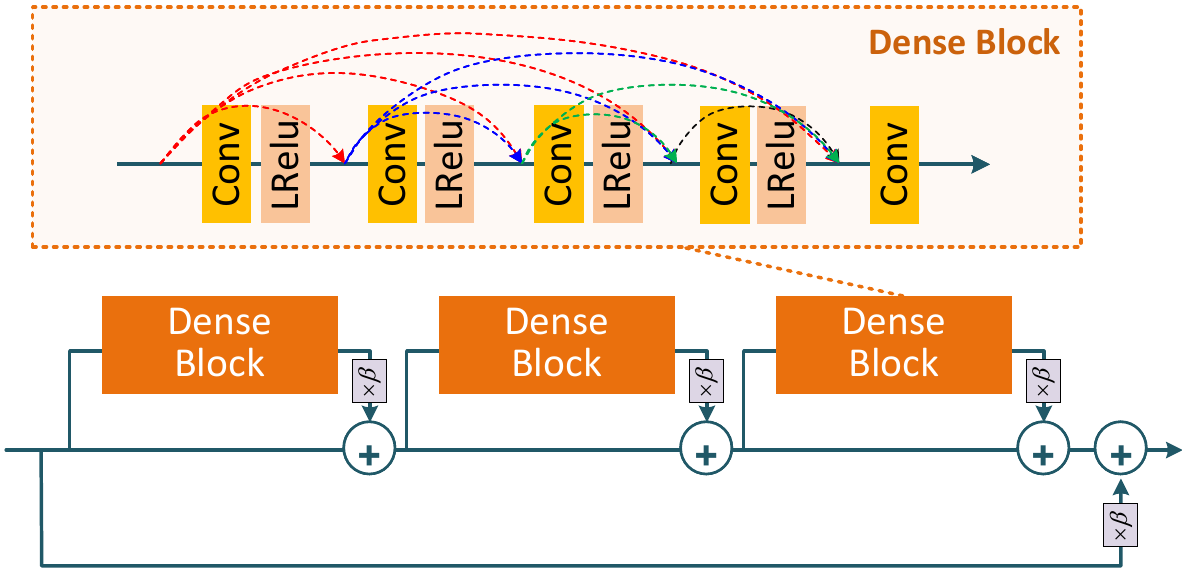}
    \caption{Architecture of the RRDB.}
    \label{fig:rrdb}
\end{figure}

We encode the input LR image as a condition of the transition probability in the reverse process. The output of the LR encoder (i.e., LR encoding) is denoted as $h_\theta$. The proposed framework can use any differentiable network structure to implement LR feature encoding. In this study, we adopt residual in residual dense block (RRDB) \cite{wang2018esrgan} as the basic feature extraction unit, which exhibits superior performance on previous SR tasks. The LR encoder, as shown in Fig.\,\ref{fig:flowchart}, contains multiple sequentially connected RRDBs and a global residual connection. The RRDB combines long-short residual connections, including multiple dense connected blocks (Dense Block). The architecture of RRDB is shown in Fig.\,\ref{fig:rrdb}.

The operation of the LR encoder is denoted as $f_\theta$, then $\mathbf{u} = f_\theta(\mathbf{X}_{\mathrm{LR}})$. In order for $f_\theta$ to effectively capture the content information in LR images, we add an upsampling layer after the LR encoder to calculate the distance between the upsampled result and the HR image. Let $f_\theta^\uparrow(\mathbf{X}_{\mathrm{LR}})$ denote the upsampled result; we calculate the $L_1$ loss between $f_\theta^\uparrow(\mathbf{X}_ {\mathrm{LR}})$ and the HR image $\mathbf{X}_{\mathrm{HR}}$ as a supervision of the LR encoder:
\begin{equation}
    \label{eq:c2fPkb}
    \mathcal L_{\mathrm{encoder}} = \mathbf E_{\mathbf{X}_{\mathrm{LR}},\mathbf{X}_{\mathrm{HR}}}\|f_\theta^\uparrow(\mathbf{X}_{\mathrm{LR}}) - \mathbf{X}_{\mathrm{HR}}\|_1.
\end{equation}

\subsection{Degradation Representation Model}
\label{sec:degradation_representation}

Degradation representation learning aims to extract discriminative degradation information from LR images. Due to the lack of degradation labels of LR images, we adopt an unsupervised contrastive learning strategy \cite{he2020momentum}. Inspired by the literature \cite{wang2021unsupervised}, we assume that image patches from the same LR image have the same degradation kernel, while degradation kernels of different LR images are different.

First, we randomly crop two image patches from an input LR image, and label one of them as a query sample and the other as a positive sample. We then crop two image patches from another LR image and label them as negative samples. Second, we employ a six-layer CNN, named degradation representation encoder, to encode the query sample, positive sample, and negative samples, and add a global average pooling (GAP) operation to obtain the corresponding degradation representation vectors, denoted as $\mathbf{v}, \mathbf{v}^+, \mathbf{v}_1^-, \mathbf{v}_2^-$, respectively. Third,  as suggested by \cite{chen2020improved}, the degradation representation vectors of these image patches are send into a three-layer MLP, and the outputs are denoted as $\mathbf{w}, \mathbf{w}^+ , \mathbf{w}_1^-, \mathbf{w}_2^-$. To make the degradation representation discriminative, $\mathbf{w}$ should be as similar as possible to $\mathbf{w}^+$ and dissimilar to $\mathbf{w}_i^-$. These similarities can be measured by \eqref{eq:jQBFS7}:
\begin{equation}
    \label{eq:jQBFS7}
    \mathcal L_{\mathbf{w}}=-\log \frac{\exp \left(\mathbf{w} \cdot \mathbf{w}^{+} / \tau\right)}{\exp(\mathbf{w} \cdot \mathbf{w}_{1}^{-} / \tau) + \exp(\mathbf{w} \cdot \mathbf{w}_{2}^{-} / \tau)},
\end{equation}
where $\tau$ is the temperature hyperparameter, and ``\,$\cdot$\,'' represents the inner product of the vectors. Studies \cite{chen2020improved,chen2020simple,park2020contrastive} have shown that constructing a large-scale set of negative samples can improve the performance of contrastive learning. Thus, during the training phase, we use a queue to store degradation representation vectors of a large number of training samples. In each iteration, the degradation representation vectors of the current batch are enqueued. If the queue is full, the earliest enqueued vectors are dequeued.

Therefore, the loss function of the degradation representation model is defined as:
\begin{equation}
    \label{eq:96jI4M}
    \mathcal L_{\mathrm{degrad}}=-\mathbf{E}_{\mathbf{w},\mathbf{w}^{+}}\log \frac{\exp \left(\mathbf{w} \cdot \mathbf{w}^{+} / \tau\right)}{\sum_{i=1}^{N_{q}}\exp(\mathbf{w} \cdot \mathbf{w}_{ q}^{(i)} / \tau)},
\end{equation}
where $N_{q}$ denotes the queue capacity, and $\mathbf{w}_{q}^{(i)}$ denotes the $i$th negative sample in the queue. The degradation representation encoder consists of six consecutive $3\times 3$ convolutional layers, and the number of output channels of these layers are $64$, $64$, $128$, $128$, $256$, $256$, respectively. The third and fifth layers are with a stride of $2$ to reduce the spatial resolution of feature maps, and the rest of the layers are with a stride of $1$. We add a batch normalization layer after each convolutional layer, and employ LeakyReLU as the activation function. Finally, the degradation representation vectors are obtained by the GAP operation.

\begin{algorithm}[t]
    \renewcommand{\algorithmicrequire}{\textbf{Input:}}
    \caption{Training}
    \label{alg:training}
    \begin{algorithmic}[1]
        \REQUIRE Total steps $T$ of the diffusion process, training samples $\{(\mathbf{X}_{\mathrm{LR}}, \mathbf{X}_{\mathrm{LR}}^+, \mathbf{X}_{\mathrm{HR}})\}$.
        \STATE{Initialize $h_\theta$, $f_\theta$, and $g_\theta$.}
        \REPEAT
        \STATE{Randomly sample $(\mathbf{X}_{\mathrm{LR}}, \mathbf{X}_{\mathrm{LR}}^+, \mathbf{X}_{\mathrm{HR}})$}
        \STATE{Compute the LR encoding $\mathbf u=f_\theta(\mathbf{X}_{\mathrm{LR}})$ and loss $\mathcal{L}_{ \mathrm{encoder}}$ by \eqref{eq:c2fPkb}.}
        \STATE{Compute the degradation representation vectors $\mathbf v=g_\theta(\mathbf{X}_{\mathrm{LR}}), \mathbf v^+=g_\theta(\mathbf{X}_{\mathrm{LR }^+})$, and loss $\mathcal{L}_{\mathrm{degrad}}$ by \eqref{eq:96jI4M}.}
        \STATE{Randomly sample $\bm\epsilon \sim \mathcal{N}(\bm 0, \bm I)$, $t\sim U(\{1, \cdots, T\})$.}
        \STATE{Substitute $(\mathbf{X}_{\mathrm{LR}}, \mathbf{X}_{\mathrm{HR}}, \mathbf u, \mathbf v, t, \bm\epsilon)$ into \eqref{eq:hWJnsO}, and then calculate the loss of the denoising model $\mathcal{L}_{\mathrm{SNF}}$.}
        \STATE{Perform a gradient descent step:
            \begin{equation*}
                \nabla_\theta \mathcal{L} = \nabla_\theta \big(\mathcal{L}_{\mathrm{SNF}} + \mathcal{L}_{\mathrm{encoder}} + \mathcal{L}_{\mathrm{degrad}}\big)
            \end{equation*}
        }
        \UNTIL{converged}
    \end{algorithmic}
\end{algorithm}

\subsection{Training}
\label{sec:training}

The training dataset consists of triples $(\mathbf{X}_{\mathrm{LR}}, \mathbf{X}_{\mathrm{LR}}^+, \mathbf{X}_{\mathrm{HR }})$, where $(\mathbf{X}_{\mathrm{LR}}, \mathbf{X}_{\mathrm{HR}})$ are paired LR-HR image patches; $\mathbf{X}_{\mathrm{LR}}^+$ and $\mathbf{X}_{\mathrm{LR}}$ are patches cropped from the same LR image with the same degradation kernel and size for degradation representation learning.

Algorithm\,\ref{alg:training} summarizes the training process. First, we randomly initialize the denoising model $h_\theta$, the LR encoder $f_\theta$ and the degradation representation model $g_\theta$. Second, we sample a batch of training images, and calculate the LR encoding $\mathbf u=f_\theta(\mathbf{X}_{\mathrm{LR}})$ and the loss of LR encoder $\mathcal{L}_{\mathrm{encoder}}$. Then, we calculate the degradation representation vectors $\mathbf{v} = g_{\theta}(\mathbf{X}_{\mathrm{LR}})$, $\mathbf{v}^+ = g_{\theta }(\mathbf{X}_{\mathrm{LR}^+})$, and the loss of the degradation representation model $\mathcal{L}_{\mathrm{degrad}}$. Finally, we randomly sample the time step $t\in\{1, \cdots, T\}$ and Gaussian noise $\bm\epsilon$, and calculate the loss of the denoising model $\mathcal{L}_{\mathrm{SNF}}$. We jointly optimize the denoising model, the LR encoder, and the degradation representation model, so all learnable parameters can be updated by computing the gradient of \eqref{eq:HyadNj}.
\begin{equation}
    \label{eq:HyadNj}
    \mathcal{L} = \mathcal{L}_{\mathrm{SNF}} + \mathcal{L}_{\mathrm{encoder}} + \mathcal{L}_{\mathrm{degrad}}.
\end{equation}

\subsection{Inference}
\label{sec:inference}

\begin{algorithm}[t]
    \renewcommand{\algorithmicrequire}{\textbf{Input:}}
    \renewcommand{\algorithmicensure}{\textbf{Output:}}
    \caption{Inference}
    \label{algo:inference}
    \begin{algorithmic}[1]
        \REQUIRE{LR image $\mathbf{X}_{\mathrm{LR}}$, sampling interval $\gamma$}
        \ENSURE{SR result $\mathbf{X}_{\mathrm{SR}}$}
        \STATE{Sampling at intervals of $\gamma$ from $\{T, T-1, \cdots, 0\}$ to obtain $\{\tau_M=T, \tau_{M-1}, \cdots,\tau_1, \tau_0=0\}$}
        \STATE{Compute the LR encoding $\mathbf u=f_\theta(\mathbf{X}_{\mathrm{LR}})$ and the degradation representation vector $\mathbf v=g_\theta(\mathbf{X} _{\mathrm{LR}})$}
        \STATE{Random sample $\mathbf{X}_T \sim \mathcal{N}(\bm 0, \bm I)$}
        \FOR{$i=M, M-1, \cdots, 2$}
        \STATE{Sample $\mathbf{z} \sim \mathcal{N}(\bm 0, \bm I)$}
        \STATE{Predict the ``clean'' image $\hat{\mathbf{X}}_0 = h_\theta(\mathbf{X}_{\tau_i}, \tau_i, \mathbf{u}, \mathbf{v})$}
        \STATE{Compute the predicted noise,
            \begin{equation*}
                \hat{\bm\epsilon} = {\left(\mathbf{X}_{\tau_i}-\sqrt{\bar{\alpha}_{\tau_i}}\,\hat{\mathbf{X}}_0\right)}/{\left(\sqrt{1-\bar{\alpha}_{\tau_i}}\right)}
            \end{equation*}
        }
        \STATE{
            Perform an update step by \eqref{eq:gs8h},
            \begin{equation*}
                \mathbf{X}_{\tau_{i-1}}\leftarrow\sqrt{\bar{\alpha}_{\tau_{i-1}}} \hat{ \mathbf{X}}_0
                + \sqrt{1-\bar{\alpha}_{\tau_{i-1}}-\sigma_{\tau_{i}}^{2}} \hat{\bm\epsilon}
                +\sigma_{\tau_{i}}\mathbf{z}
            \end{equation*}
        }
        \ENDFOR
        \STATE{Compute $\hat{\mathbf{X}}_{0}=h_\theta(\mathbf{X}_{\tau_1}, \tau_1, \mathbf{u}, \mathbf{v})=:\mathbf{X}_{\mathrm{SR}}$ by \eqref{eq:jSWGEw}}
    \end{algorithmic}
\end{algorithm}

In the inference phase, we first calculate the LR encoding $\mathbf{u}$ and the degradation representation vector $\mathbf{v}$ of the input LR image. Then, with $\mathbf{u}$ and $\mathbf{v}$ as conditions, we start from a Gaussian noise $\mathbf{X}_T$, iteratively predict and remove the noise added by the forward process, and finally generate the SR result. Here, the LR encoding and the degradation representation vector only need to be computed once, while the denoising model needs to be performed repeatedly according to the number of iterations. Therefore, when the number of steps $T$ of the reverse process is large, the inference of the model will take a long time. To improve the inference efficiency, we adopt the accelerated sampling strategy proposed in \cite{song2020denoising}, which selects a subset of the reverse process time series to reduce the number of iterations. This strategy can greatly improve the inference speed without compromising the quality of the SR results. Specifically, the reverse process starts from the $T$th step, sampling every $\gamma$ steps (assuming $\gamma$ can divide $T$) to obtain a new reverse process sampling path:
\begin{equation}
    T \rightarrow (T-\gamma) \rightarrow (T-2\gamma) \rightarrow \cdots \rightarrow 0.
\end{equation}
Let $\{\tau_0, \tau_1, \cdots, \tau_M\}$ denote the new reverse process sampling path, where $\tau_0 = 0$ and $M=T/\gamma$. In this study, we simulate a trajectory of the reverse process on the simplified sampling path to generate SR results. When $i > 1$, samples can be generated iteratively:
\begin{equation}
    \label{eq:gs8h}
    \begin{aligned}
         & \mathbf{X}_{\tau_{i-1}}=  \sqrt{\bar{\alpha}_{\tau_{i-1}}}
        \underbrace{h_\theta(\mathbf{X}_{\tau_i}, \tau_i, \mathbf{u}, \mathbf{v})}_{\text{``predicted clean image $\mathbf X_0$''}}                                                                                     \\
         & + \sqrt{1-\bar{\alpha}_{\tau_{i-1}}-\sigma_{\tau_{i}}^{2}}
        \underbrace{\frac{\mathbf{X}_{\tau_i}-\sqrt{\bar{\alpha}_{\tau_i}}\,h_\theta(\mathbf{X}_{\tau_i}, \tau_i, \mathbf{u}, \mathbf{v})}{\sqrt{1-\bar{\alpha}_{\tau_i}}}}_{\text{``predicted noise $\bm\epsilon $''}} \\
         & +  \underbrace{\sigma_{\tau_{i}} \mathbf{z}_{\tau_{i}}}_{\text{``random Noise''}},
    \end{aligned}
\end{equation}
where $\mathbf z_{\tau_i}\sim\mathcal N(\mathbf 0, \bm I)$, $\sigma_{\tau_{i}}=\eta \sqrt{\frac{1-\bar {\alpha}_{\tau_{i-1}}}{1-\bar{\alpha}_{\tau_{i}}} \beta_{\tau_{i}}}$, and $\eta$ is the temperature coefficient that controls the variance. When $i = 1$, the last step of the reverse process is reached.
\begin{equation}
    \label{eq:jSWGEw}
    \mathbf{X}_{0} = h_\theta(\mathbf{X}_{\tau_1}, \tau_1, \mathbf{u}, \mathbf{v}).
\end{equation}
If $\eta = 0$, the sample generation process is no longer stochastic, and the model degrades into a deterministic mapping from $\mathbf{X}_T$ to $\mathbf{X}_0$. The diversity of samples will increase as $\eta\ (< 1)$ increases. Algorithm \ref{algo:inference} demonstrates the inference phase.

\section{Experimental Results}
\label{sec:experiments}

\begin{table*}[ht]
    \centering
    \renewcommand\arraystretch{1.4}
    \caption{Quantitative Comparison of PSNR(\si{\dB}), FID and LPIPS for Noise-Free Isotropic Degradations on the GeoEye-1 and GoogleEarth Datasets. \textbf{Bold} Represents the Best Results.}
    \label{tab:main-iso}
    \setlength{\tabcolsep}{1.6mm}{
        \begin{tabular}{r|l|ccc|ccc|ccc|ccc}
            \hline
            \hline
            \multirow{2}{6em}{Datasets} & Kernel width $\sigma$           & \multicolumn{3}{c|}{0}                            & \multicolumn{3}{c|}{1.2}                          & \multicolumn{3}{c|}{2.4}                          & \multicolumn{3}{c}{3.6} \bigstrut \\
            \cline{2-14}
             & Metrics                         & PSNR           & FID    & LPIPS  & PSNR           & FID    & LPIPS  & PSNR           & FID    & LPIPS  & PSNR           & FID    & LPIPS \bigstrut \\
            \hline\hline
            \multirow{8}{6em}{GeoEye-1} & Bicubic                         & 21.26          & 171.71 & 0.6259 & 20.89          & 187.75 & 0.6590 & 20.06          & 235.02 & 0.7400 & 19.36          & 279.37 & 0.8171  \bigstrut[t] \\
             & SAN \cite{dai2019second}        & \textbf{22.64} & 143.36 & 0.3696 & 22.21          & 165.09 & 0.4157 & 20.49          & 211.02 & 0.6208 & 19.48          & 269.54 & 0.7749 \\
             & ESRGAN \cite{wang2018esrgan}    & 19.70          & 86.20  & 0.1620 & 20.30          & 105.77 & 0.1919 & 20.06          & 216.04 & 0.3883 & 19.22          & 281.31 & 0.6116 \\
             & ZSSR \cite{shocher2018zero}     & 21.53          & 168.52 & 0.5861 & 21.32          & 171.59 & 0.6016 & 20.36          & 216.40 & 0.6878 & 19.49          & 267.67 & 0.7819 \\
             & DASR \cite{wang2021unsupervised} & 22.39          & 147.04 & 0.4398 & \textbf{22.41} & 148.02 & 0.4428 & \textbf{22.37} & 153.89 & 0.4501 & \textbf{21.87} & 191.14 & 0.4809 \\
             & Real-ESRGAN \cite{wang2021real} & 18.00          & 141.77 & 0.3001 & 17.88          & 143.32 & 0.3049 & 17.58          & 150.63 & 0.3156 & 17.22          & 159.73 & 0.3303 \\
             & BSRGAN \cite{zhang2021designing} & 18.23          & 169.40 & 0.2810 & 18.20          & 169.60 & 0.2845 & 17.99          & 167.30 & 0.2957 & 17.60          & 171.84 & 0.3155 \\
             & BlindSRSNF (Ours)               & 20.80          & \textbf{68.96} & \textbf{0.1459} & 20.87          & \textbf{67.81} & \textbf{0.1459} & 20.82          & \textbf{66.65} & \textbf{0.1485} & 20.24          & \textbf{80.78} & \textbf{0.1703} \bigstrut[b] \\
            \hline
            \multirow{8}{6em}{GoogleEarth} & Bicubic                         & 23.74          & 145.94 & 0.6256 & 23.28          & 150.08 & 0.6572 & 22.22          & 182.61 & 0.7342 & 21.28          & 223.27 & 0.8187  \bigstrut[t] \\
             & SAN \cite{dai2019second}        & 25.06          & 115.38 & 0.4296 & 24.67          & 123.88 & 0.4476 & 22.76          & 171.65 & 0.5854 & 21.43          & 209.77 & 0.7529 \\
             & ESRGAN \cite{wang2018esrgan}    & 22.40          & 101.35 & 0.3103 & 22.86          & 118.89 & 0.3123 & 22.22          & 152.25 & 0.4695 & 21.13          & 199.34 & 0.6444 \\
             & ZSSR \cite{shocher2018zero}     & 24.03          & 161.41 & 0.5394 & 23.81          & 156.37 & 0.5630 & 22.61          & 183.93 & 0.6575 & 21.46          & 218.04 & 0.7715 \\
             & DASR \cite{wang2021unsupervised} & \textbf{25.14} & 132.16 & 0.4708 & \textbf{25.14} & 134.75 & 0.4712 & \textbf{25.07} & 133.25 & 0.4777 & \textbf{24.50} & 181.73 & 0.5017 \\
             & Real-ESRGAN \cite{wang2021real} & 19.45          & 162.62 & 0.3286 & 19.34          & 160.66 & 0.3348 & 19.02          & 172.25 & 0.3487 & 18.62          & 180.67 & 0.3653 \\
             & BSRGAN \cite{zhang2021designing} & 21.32          & 136.59 & 0.2840 & 21.27          & 137.76 & 0.2875 & 21.03          & 138.78 & 0.2966 & 20.57          & 147.27 & 0.3138 \\
             & BlindSRSNF (Ours)               & 23.28          & \textbf{70.01} & \textbf{0.2090} & 23.30          & \textbf{69.06} & \textbf{0.2103} & 23.26          & \textbf{72.52} & \textbf{0.2145} & 22.71          & \textbf{80.36} & \textbf{0.2373} \bigstrut[b] \\
            \hline
            \hline
        \end{tabular}%
    }
\end{table*}

In this section, we first introduce the datasets and evaluation metrics. Then, the model settings and training details are presented. Finally, we demonstrate the effectiveness of the proposed method using various degradation kernels and real-world RSIs.

\subsection{Datasets and Metrics}
We use RSIs provided by GeoEye-1 satellite and GoogleEarth to verify the effectiveness of our proposal. The GeoEye-1 dataset contains $130$ multispectral images with a resolution of $\SI{0.41}{\metre}$ and a size of $512\times 512$, of which $115$ are used for training, and the remaining $15$ are used for testing. The GoogleEarth dataset contains $239$ optical RSIs with a resolution of $\SI{1}{\metre}$ and a size of $512\times 512$, of which $224$ are used for training and the remaining $15$ are used for testing. In our experiments, the training set contains a total of $339$ RSIs from the above two sources.

The proposed BlindSRSNF aims to generate more realistic SR results for real-world RSIs. Besides the common used peak signal-to-noise ratio (PSNR), we also introduce two objective metrics FID \cite{heusel2017gans} and LPIPS \cite{zhang2018the} to better measure the visual quality of SR results.
FID can better evaluate the quality and diversity of images generated by the model, and LPIPS can obtain evaluations that are almost consistent with human visual perception \cite{lugmayr2020ntire}.

\subsection{Implementation Details}
In the SNF, the number of diffusion steps $T$ is set to $1000$, and the noise variance is reduced from $\beta_1 = 2\times 10^{-2}$ to $\beta_{T} = 1\times 10^{-4}$ by using the setting in \cite{nichol2021improved}. The basic channel number $c$ of the convolutional layers in the denoising network is set to $64$. The sampling interval $\gamma$ is set to $50$, and the temperature coefficient $\eta$ is fixed to $1$. The setting of the sampling interval $\gamma$ is discussed in Sec.\,\ref{sec:ablation}. The number of RRDBs in the LR encoder is set to $23$, and the number of channels in each RRDB is set to $64$. We follow \cite{wang2021unsupervised} to generate LR images using two degradation models. The first model degrades the images using isotropy Gaussian blur kernels without adding noise. The size of kernels is fixed at $21\times 21$, and the kernel width obeys a uniform distribution, $\sigma\sim U(0.2, 4.0)$. The second model degrades the images using anisotropy Gaussian blur kernels and then adds additive white Gaussian noise. The size of the blur kernel is fixed at $21\times 21$. The covariance matrix of the blur kernel is determined by two random eigenvalues $\lambda_1, \lambda_2 \sim U(0.2, 4)$ and a random rotation angle $\theta\sim U(0, \pi)$.
The noise level various randomly from $0$ to $25$.

In the training phase, we randomly crop the degraded LR images into patches with a size of $64\times 64$ as input, and randomly flip vertically or horizontally, and random rotate $90^\circ$ for data augmentation. The proposed method is implemented based on the PyTorch framework and trained on an NVIDIA GeForce RTX 3090 GPU.

\subsection{Comparison on Noise-free Degradations with Isotropy Gaussian Kernels}

\begin{figure*}[t]
    \centering
    \includegraphics[width=\textwidth]{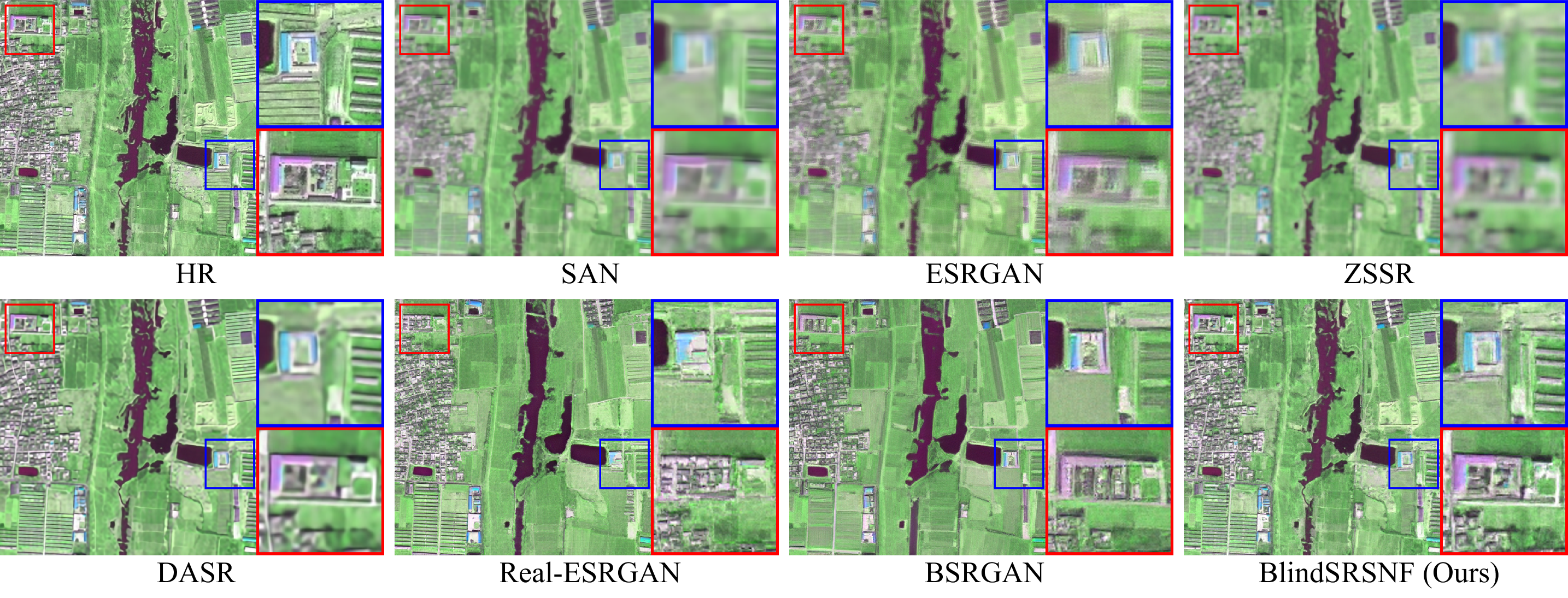}
    \\\vspace{0.5em}
    \includegraphics[width=\textwidth]{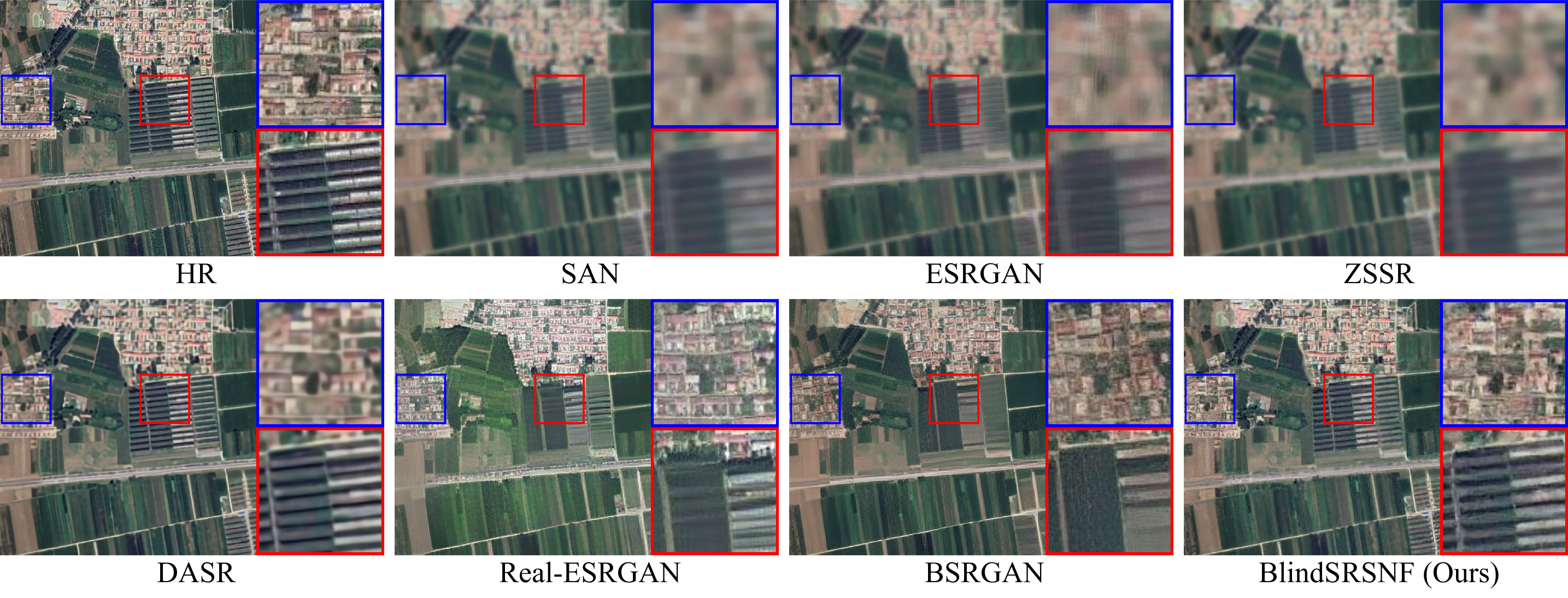}
    \caption{Visual comparison for noise-free isotropic degradation kernels. The top two rows are from the GeoEye-1 dataset with kernel width $\sigma = 2.4$ and the bottom two rows are from the GoogleEarth dataset with kernel width $\sigma = 3.6$.}
    \label{fig:main-iso}
\end{figure*}

We take bicubic interpolation as the baseline and compare our proposed BlindSRSNF with six SOTA SR algorithms, including SAN \cite{dai2019second}, ESRGAN \cite{wang2018esrgan}, ZSSR \cite{shocher2018zero}, DASR \cite{wang2021unsupervised}, Real-ESRGAN \cite{wang2021real}, and BSRGAN \cite{zhang2021designing}.

SAN is an excellent SR method optimized by pixel-level loss, and ESRGAN is a perceptual loss-optimized algorithm with good visual results. The above two algorithms assume that the degradation model is bicubic. ZSSR is an unsupervised blind SR algorithm that trains a small CNN to the specific test image during the inference phase. DASR is a blind SR algorithm based on contrastive learning, optimized by pixel-level loss. Real-ESRGAN and BSRGAN are SOTA GAN-based blind SR algorithms with more visual pleasing results than methods optimized by pixel-level losses. For a fair comparison, all competitive algorithms are retrained on the same training datasets by using their public codes.

Table \ref{tab:main-iso} shows the comparison of objective metrics with kernel widths of $0$, $1.2$, $2.4$ and $3.6$ on the GeoEye-1 and GoogleEarth datasets, respectively. When $\sigma=0$, the degradation model is reduced to the ordinary bicubic degradation. Since SAN and ESRGAN are trained based on the bicubic assumption, better results can be obtained when $\sigma=0$. However, SAN and ESRGAN are difficult to deal with other complex degradations. ZSSR and DASR are optimized by pixel-level losses, which is equivalent to directly optimizing PSNR, so higher PSNR values can be obtained. However, they perform poorly on the perceptual metrics, FID and LPIPS. Real-ESRGAN and BSRGAN achieve better FID and LPIPS scores but significantly lower PSNR than DASR. The proposed BlindSRSNF is optimized by the negative log-likelihood, which significantly improves the FID and LPIPS metrics, achieving the best performance on all degradations. Furthermore, the PSNR of BlindSRSNF significantly outperforms GAN-based methods by up to $\SI{4.24}{dB}$. The results show that out proposal can generate higher-quality SR results while significantly reducing the spatial distortions.

Fig.\,\ref{fig:main-iso} shows the visual comparison on noise-free isotropic kernels. It can be seen that SAN and ESRGAN trained based on the bicubic assumption are difficult to remove blur effectively. DASR can effectively deal with various degradations, but due to its optimization goal, the generated results are too smooth and lack texture details. Although the results of Real-ESRGAN are visually realistic, their content deviates greatly from real HR image, and even changes the category of ground objects. BlindSRSNF obtains SR results with the best visual perceptual quality. The great visual quality are attributed to the stochastic normalized flow, which allow the model can explicitly learn the probability distribution in HR space through maximum likelihood estimation. Besides, since BlindSRSNF contains an LR encoder, the texture details of the SR results are in good agreement with real HR images. The integrated degradation representation model and the conditional probability transition paradigm also make it possible to adapt the BlindSRSNF to multiple degradations.

\subsection{Comparison on General Degradations with Anisotropy Gaussian Kernels}

\begin{table*}[t]
    \centering
    \renewcommand\arraystretch{1.4}
    \caption{Quantitative Comparison of PSNR(\si{\dB}), FID and LPIPS for Anisotropic Degradation Kernel on the GeoEye-1 Dataset. \\ \textbf{Bold} Represents the Best Results.}
    \label{tab:main-aniso}
    \setlength{\tabcolsep}{2mm}{
        \begin{tabular}{l|c|ccc|ccc|ccc|ccc}
            \hline
            \hline
            \multirow{3}{*}[1em]{Methods}   & \multirow{3}{*}[1em]{\makecell{Noise \\Levels}} & \multicolumn{3}{c|}{\begin{minipage}[b]{40pt}\vspace{5pt} \centering\includegraphics[width=30pt]{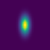} \end{minipage}} & \multicolumn{3}{c|}{\begin{minipage}[b]{40pt}\vspace{5pt} \centering\includegraphics[width=30pt]{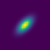}  \end{minipage}} & \multicolumn{3}{c|}{\begin{minipage}[b]{40pt}\vspace{5pt} \centering\includegraphics[width=30pt]{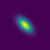} \end{minipage}} & \multicolumn{3}{c}{\begin{minipage}[b]{40pt}\vspace{5pt} \centering\includegraphics[width=30pt]{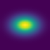}  \end{minipage}} \bigstrut[t] \\
                                            &                        & \multicolumn{3}{c|}{$[\lambda_1, \lambda_2, \theta] = [1.2, 2.4, 0]$} & \multicolumn{3}{c|}{$[\lambda_1, \lambda_2, \theta] = [1.2, 2.4, \pi/4]$} & \multicolumn{3}{c|}{$[\lambda_1, \lambda_2, \theta] = [2.4, 1.2, \pi/4]$} & \multicolumn{3}{c}{$[\lambda_1, \lambda_2, \theta] = [3.6, 2.4, 0]$} \bigstrut[b] \\\cline{3-14}
                                            &                        & PSNR           & FID            & LPIPS  & PSNR           & FID    & LPIPS  & PSNR           & FID    & LPIPS  & PSNR           & FID            & LPIPS \bigstrut \\
            \hline
            \hline
            Bicubic                         & \multirow{8}[2]{*}{0}  & 20.38          & 205.88         & 0.6848 & 20.45          & 219.45 & 0.6982 & 20.41          & 212.19 & 0.6913 & 19.71          & 264.68         & 0.7904  \bigstrut[t] \\
            SAN \cite{dai2019second}        &                        & 21.01          & 183.70         & 0.5248 & 21.23          & 197.02 & 0.5144 & 21.13          & 205.26 & 0.5174 & 19.98          & 253.60         & 0.7099 \\
            ESRGAN \cite{wang2018esrgan}    &                        & 20.30          & 175.04         & 0.2990 & 20.37          & 152.61 & 0.2749 & 20.32          & 162.05 & 0.2831 & 19.64          & 255.04         & 0.5065 \\
            ZSSR \cite{shocher2018zero}     &                        & 20.73          & 189.72         & 0.6249 & 20.84          & 211.07 & 0.6412 & 20.80          & 197.62 & 0.6304 & 19.93          & 249.91         & 0.7485 \\
            DASR \cite{wang2021unsupervised} &                        & \textbf{22.23} & 153.53         & 0.4865 & \textbf{21.91} & 188.08 & 0.4955 & \textbf{21.86} & 178.72 & 0.4922 & \textbf{21.99} & 173.81         & 0.5074 \\
            Real-ESRGAN \cite{wang2021real} &                        & 17.71          & 151.46         & 0.3112 & 17.73          & 149.24 & 0.3096 & 17.71          & 150.70 & 0.3118 & 17.39          & 155.40         & 0.3240 \\
            BSRGAN \cite{zhang2021designing} &                        & 18.03          & 177.70         & 0.2953 & 18.09          & 167.54 & 0.2901 & 18.10          & 167.84 & 0.2931 & 17.86          & 176.80         & 0.3053 \\
            BlindSRSNF (Ours)               &                        & 20.34          & \textbf{78.33} & \textbf{0.1687} & 20.19          & \textbf{110.58} & \textbf{0.1923} & 20.14          & \textbf{108.34} & \textbf{0.1880} & 19.97          & \textbf{86.74} & \textbf{0.1850} \bigstrut[b] \\
            \hline
            Bicubic                         & \multirow{8}[2]{*}{5}  & 20.26          & 218.79         & 0.7231 & 20.33          & 237.80 & 0.7378 & 20.29          & 225.35 & 0.7304 & 19.61          & 268.89         & 0.7780  \bigstrut[t] \\
            SAN \cite{dai2019second}        &                        & 20.51          & 177.72         & 0.6038 & 20.63          & 199.76 & 0.5935 & 20.58          & 199.11 & 0.5977 & 19.66          & 263.03         & 0.6932 \\
            ESRGAN \cite{wang2018esrgan}    &                        & 16.78          & 298.66         & 0.5370 & 16.83          & 301.99 & 0.5115 & 16.78          & 290.00 & 0.5161 & 16.66          & 386.66         & 0.6038 \\
            ZSSR \cite{shocher2018zero}     &                        & 20.52          & 222.20         & 0.6842 & 20.61          & 233.82 & 0.7073 & 20.56          & 225.97 & 0.6983 & 19.75          & 278.85         & 0.7585 \\
            DASR \cite{wang2021unsupervised} &                        & \textbf{21.57} & 203.44         & 0.5177 & \textbf{21.54} & 211.68 & 0.5221 & \textbf{21.50} & 213.41 & 0.5211 & \textbf{20.89} & 253.19         & 0.5722 \\
            Real-ESRGAN \cite{wang2021real} &                        & 17.66          & 152.78         & 0.3214 & 17.68          & 156.64 & 0.3204 & 17.66          & 155.81 & 0.3221 & 17.33          & 160.28         & 0.3358 \\
            BSRGAN \cite{zhang2021designing} &                        & 17.96          & 180.72         & 0.3005 & 18.02          & 172.39 & 0.2950 & 18.02          & 171.68 & 0.2979 & 17.75          & 186.84         & 0.3142 \\
            BlindSRSNF (Ours)               &                        & 19.63          & \textbf{109.83} & \textbf{0.2065} & 19.68          & \textbf{116.16} & \textbf{0.2120} & 19.65          & \textbf{121.67} & \textbf{0.2108} & 18.88          & \textbf{128.41} & \textbf{0.2376} \bigstrut[b] \\
            \hline
            Bicubic                         & \multirow{8}[2]{*}{10} & 19.91          & 271.32         & 0.7802 & 19.98          & 282.74 & 0.7922 & 19.95          & 274.79 & 0.7874 & 19.31          & 321.13         & 0.8173  \bigstrut[t] \\
            SAN \cite{dai2019second}        &                        & 19.41          & 243.57         & 0.7396 & 19.50          & 255.53 & 0.7326 & 19.47          & 254.70 & 0.7378 & 18.77          & 318.97         & 0.7875 \\
            ESRGAN \cite{wang2018esrgan}    &                        & 15.73          & 326.74         & 0.6659 & 15.76          & 318.02 & 0.6451 & 15.74          & 315.30 & 0.6490 & 15.56          & 368.41         & 0.7235 \\
            ZSSR \cite{shocher2018zero}     &                        & 19.88          & 278.85         & 0.7741 & 19.99          & 288.66 & 0.7833 & 19.93          & 289.16 & 0.7835 & 19.23          & 334.64         & 0.8239 \\
            DASR \cite{wang2021unsupervised} &                        & \textbf{20.96} & 240.48         & 0.5463 & \textbf{21.05} & 238.94 & 0.5494 & \textbf{21.00} & 240.42 & 0.5489 & \textbf{20.32} & 275.06         & 0.6156 \\
            Real-ESRGAN \cite{wang2021real} &                        & 17.56          & 157.76         & 0.3420 & 17.59          & 158.60 & 0.3419 & 17.57          & 158.99 & 0.3419 & 17.23          & 170.71         & 0.3574 \\
            BSRGAN \cite{zhang2021designing} &                        & 17.78          & 189.33         & 0.3170 & 17.84          & 179.46 & 0.3123 & 17.84          & 183.72 & 0.3148 & 17.52          & 201.41         & 0.3390 \\
            BlindSRSNF (Ours)               &                        & 19.02          & \textbf{124.33} & \textbf{0.2381} & 19.11          & \textbf{129.18} & \textbf{0.2405} & 19.07          & \textbf{133.92} & \textbf{0.2395} & 18.30          & \textbf{143.42} & \textbf{0.2682} \bigstrut[b] \\
            \hline
            \hline
        \end{tabular}%
    }
\end{table*}

\begin{figure*}[htp]
    \centering
    \includegraphics[width=\textwidth]{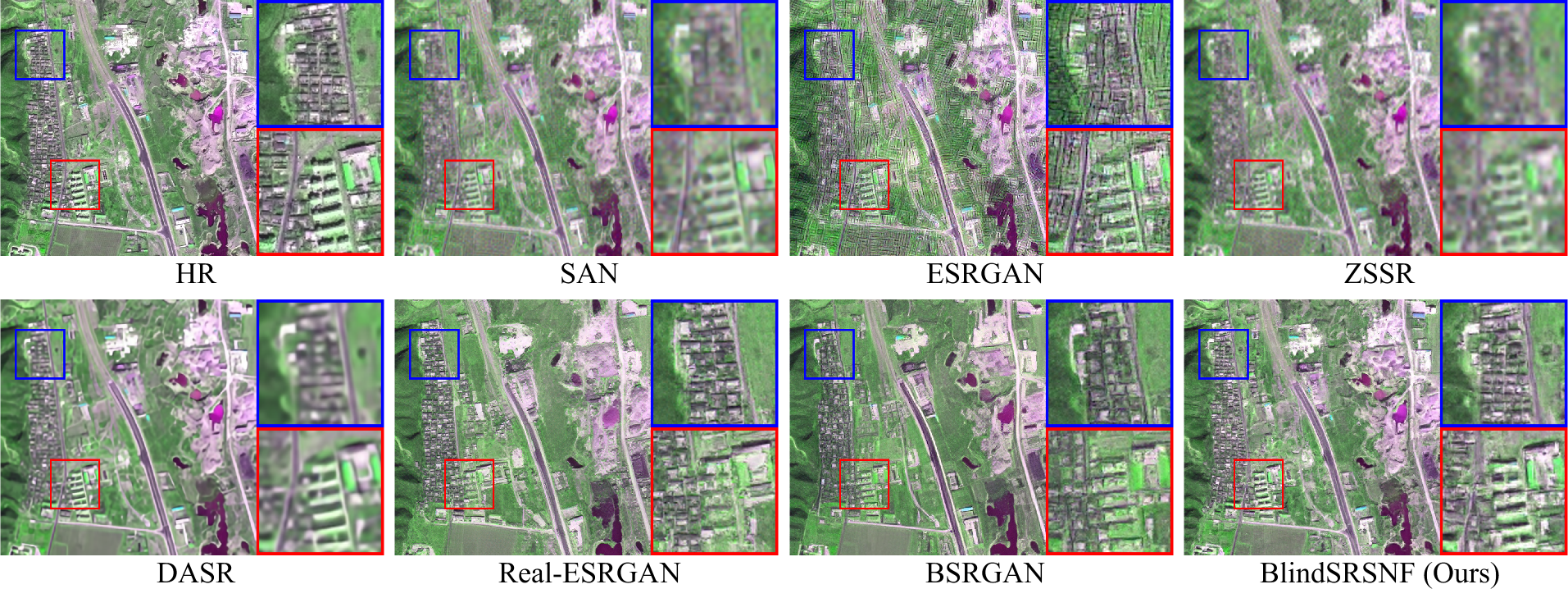}
    \\\vspace{0.5em}
    \includegraphics[width=\textwidth]{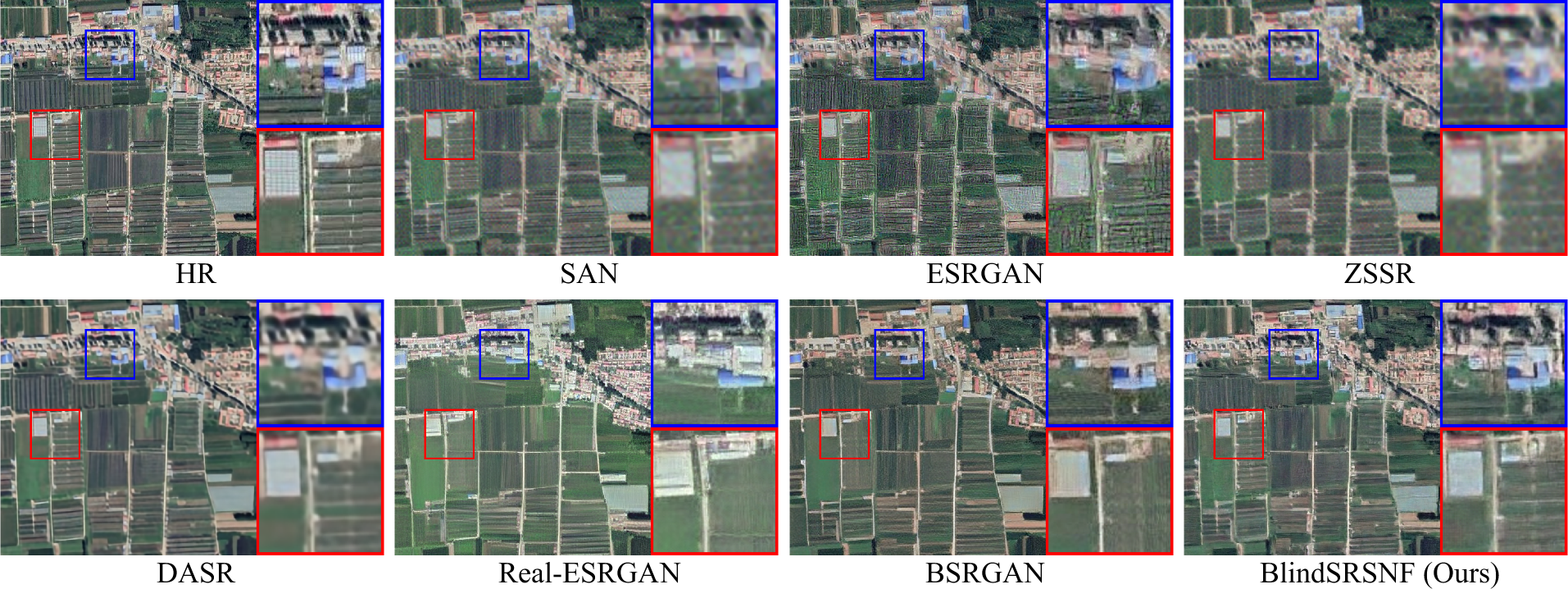}
    \\\vspace{0.5em}
    \includegraphics[width=\textwidth]{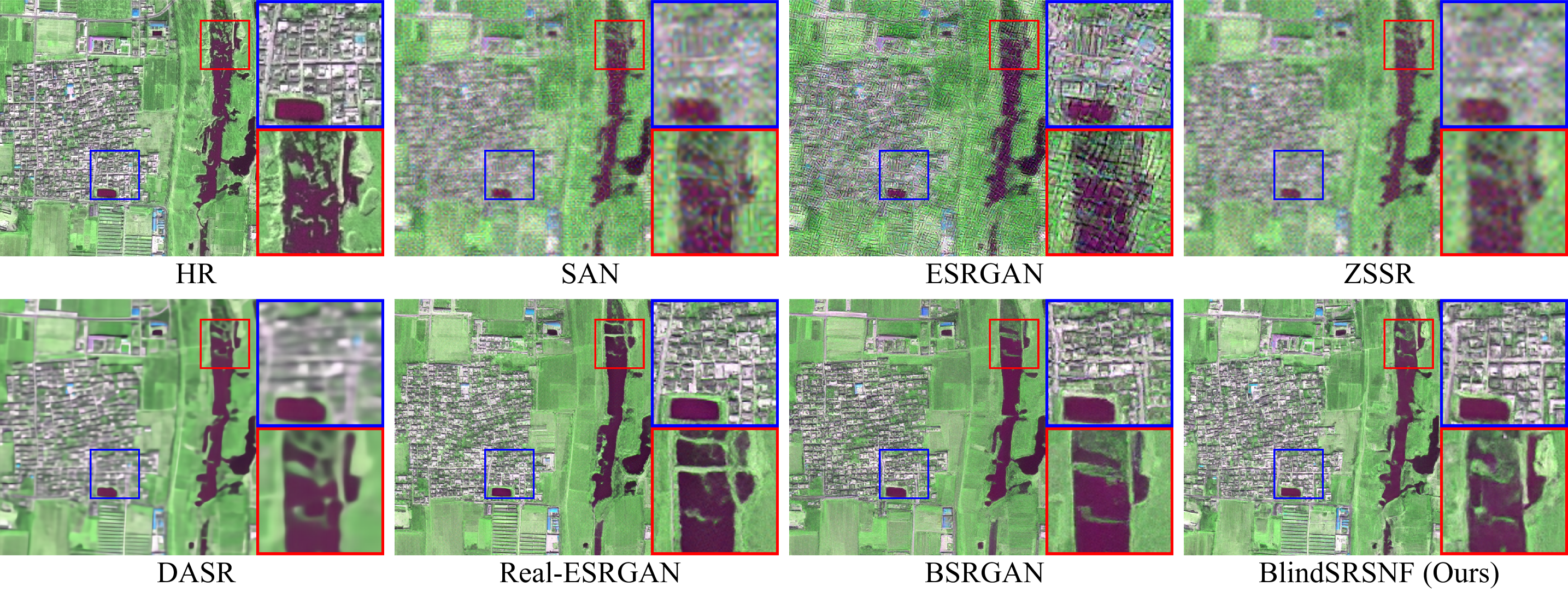}
    \caption{Visual comparison for anisotropic degradation kernels. The top two rows are from the GeoEye-1 dataset with a blur kernel of $\lambda_1=0.6, \lambda_2=2.4, \theta=0$ and a noise level of $5$. The middle two rows are from the GoogleEarth dataset with a blur kernel of $\lambda_1=2.4, \lambda_2=1.2, \theta=\pi/4$ and a noise level of $5$. The bottom two rows are from the GeoEye-1 dataset with blur a kernel of $\lambda_1=3.6, \lambda_2=2.4, \theta=0$ and a noise level of $10$.}
    \label{fig:main-aniso}
\end{figure*}

We adopt the more general degradations with anisotropic Gaussian kernels to compare the proposed BlindSRSNF with six SR algorithms, including SAN\cite{dai2019second}, ESRGAN\cite{wang2018esrgan}, ZSSR\cite{shocher2018zero}, DASR\cite{wang2021unsupervised}, Real-ESRGAN\cite{wang2021real}, BSRGAN\cite{zhang2021designing}. Table \ref{tab:main-aniso} shows the comparison of objective metrics for general degradations on the GeoEye-1 dataset. Since the blur kernels of these degradation models are more general and additional noise is added, the performance of all the competitive methods on this more difficult task degrades.

The DASR optimized by pixel-level loss achieves the highest PSNR, but performs poorly on visual perceptual metrics, FID and LPIPS. The GAN-based models significantly outperform DASR on visual perception metrics. The proposed BlindSRSNF outperforms all competitive algorithms in terms of FID and LPIPS, with an improvement of up to $48\%$ in FID and up to $46\%$ in LPIPS compared to the second-ranked Real-ESRGAN. Although the PSNR of BlindSRSNF is lower than DASR, BlindSRSNF achieves the best PSNR among all perception-optimized blind SR algorithms, and achieves a $\SI{2.31}{\dB}$ improvement over the second-ranked BSRGAN.

Fig.\,\ref{fig:main-aniso} shows the visual comparison for anisotropic degradation kernels. We randomly select four sets of blur kernel parameters and with three noise levels for demonstration. It can be seen that noise will seriously affect the SR results. Methods trained on bicubic degradations can barely recover texture details. The results of DASR are too smooth to clearly distinguish between houses and roads in residential areas. Although the results of Real-ESRGAN look clear, the texture details of real HR images are severely falsified, such as changing the layout of houses, the location of roads, and the shape of rivers. Besides, Real-ESRGAN changes the spectral information of the images in the GoogleEarth dataset, manifesting as a significant color shift. The textures generated by BSRGAN and BlindSRSNF are more realistic, but BSRGAN is not as sharp as BlindSRSNF. Combining the results in in Fig.\,\ref{fig:main-aniso} and the LPIPS scores in Table \ref{tab:main-aniso}, it can be found that the LPIPS can obtain objective evaluation consistent with the quality of human visual perception. In conclusion, the proposed BlindSRSNF can generate blind SR results with the best visual perceptual quality.

\subsection{Comparison on Real-World RSIs}

\begin{figure*}[t]
    \centering
    \includegraphics[width=\textwidth]{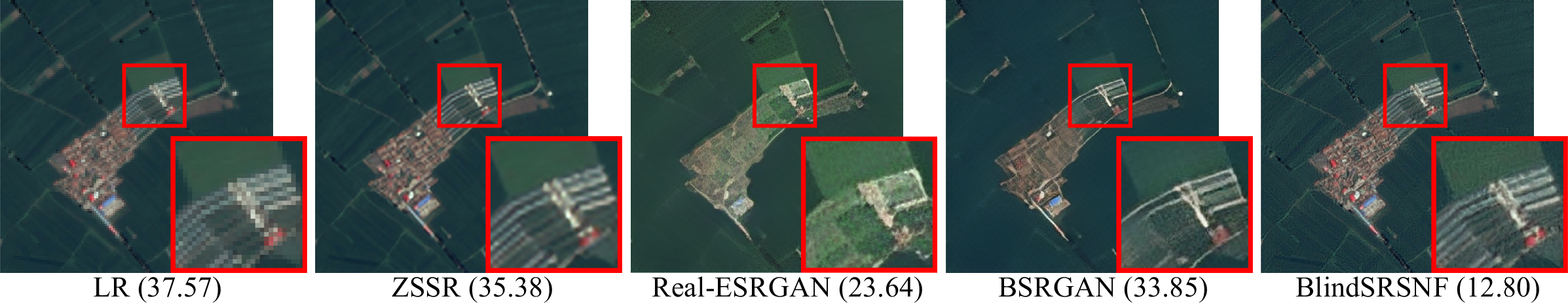}
    \includegraphics[width=\textwidth]{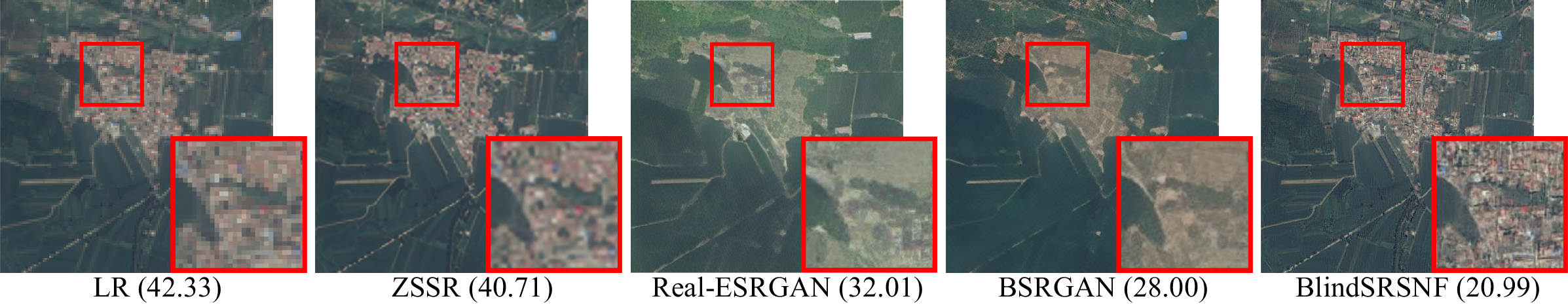}
    \includegraphics[width=\textwidth]{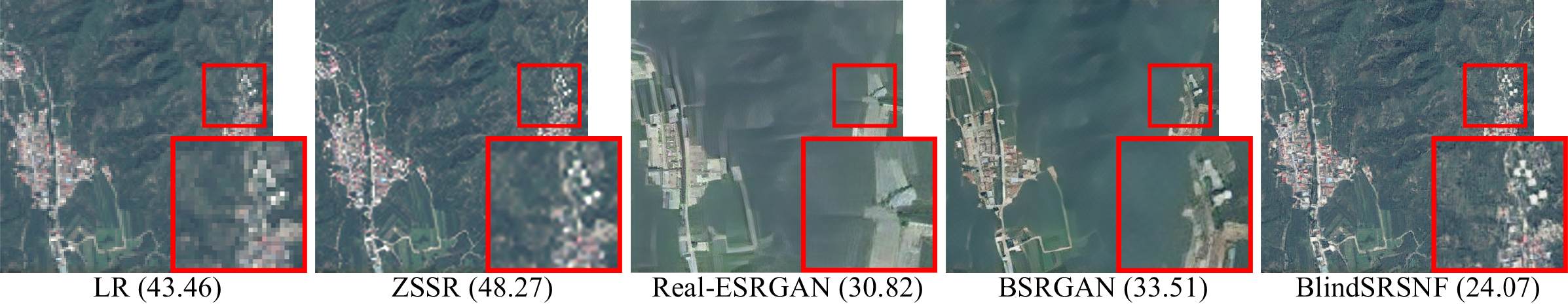}
    \caption{Visual comparison on real-world RSIs. The values in parentheses are BRISQUE scores.}
    \label{fig:main-real}
\end{figure*}

In this section, we perform SR on real-world RSIs (rather than simulated LR images) to verify the performance of the proposed method in real scenarios. LR images are from the GoogleEarth dataset, and the comparison algorithms include ZSSR \cite{shocher2018zero}, Real-ESRGAN \cite{wang2021real}, BSRGAN \cite{zhang2021designing} and the proposed BlindSRSNF. The visual results are shown in Fig.\,\ref{fig:main-real}. Due to the lack of ground truths, we adopt a blind/referenceless image spatial quality evaluator (BRISQUE) \cite{mittal2012no} to measure the quality of the SR results. It can be observed that the texture details of ZSSR are relatively blurred, and the two GAN-based algorithms have tampered with the contents such as forest and farmland areas. Although Real-ESRGAN obtain the second BRISQUE after our proposed BlindSRSNF, it suffers from severe spectral shift and produces very unrealistic texture details. In summary, the proposed BlindSRSNF can obtain more realistic and clear SR results in real scenarios.

\begin{figure*}[ht]
    \centering
    \includegraphics[width=\textwidth]{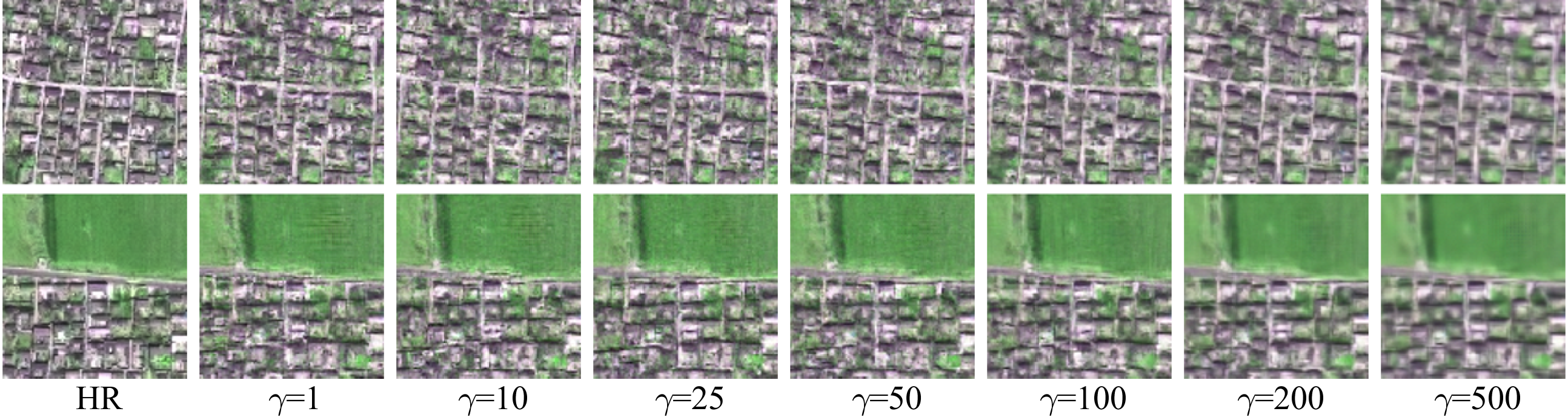}
    \caption{Comparison of visual results with different sampling intervals.}
    \label{fig:ablation-gamma}
\end{figure*}

\subsection{Ablation Studies}
\label{sec:ablation}
\begin{table*}[ht]
    \centering
    \renewcommand\arraystretch{1.4}
    \caption{Discussion of Degradation Representation Learning. Metrics are PSNR(\si{\dB}), FID and LPIPS. \textbf{Bold} Represents the Best Results.}
    \label{tab:ablation-contrast}
    \setlength{\tabcolsep}{1.8mm}{
        \begin{tabular}{c|ccc|ccc|ccc}
            \hline
            \hline
            Noise level & \multicolumn{3}{c|}{0}                            & \multicolumn{3}{c|}{5}                             & \multicolumn{3}{c}{10} \bigstrut \\
            \hline
            Metrics     & PSNR           & FID   & LPIPS  & PSNR           & FID    & LPIPS  & PSNR           & FID    & LPIPS \bigstrut \\
            \hline
            BlindSRSNF w/o degradation representation learning & 19.67          & 87.63 & 0.1885 & 18.62          & 132.92 & 0.2412 & 18.02          & 156.55 & 0.2721  \bigstrut[t] \\
            BlindSRSNF  & \textbf{19.97} & \textbf{86.74} & \textbf{0.1850} & \textbf{18.88} & \textbf{128.41} & \textbf{0.2376} & \textbf{18.30} & \textbf{143.42} & \textbf{0.2682}  \bigstrut[b] \\
            \hline
            \hline
        \end{tabular}%
    }
\end{table*}

In this section, we first verify the effectiveness of the degraded representation learning. Then, we discuss the parameter settings of the SNF model.

\subsubsection{Degradation Representation Learning}
We construct a contrastive model without the degradation representation learning module, where the degradation-aware convolutional layers in the denoising network are replaced by ordinary convolutional layers, and the contrastive loss is removed. Table \ref{tab:ablation-contrast} shows the ablation results, which were tested on the GeoEye-1 dataset. The blur kernel parameters are $\lambda_1=3.6,\lambda_2=2.4,\theta=0$ and the noise levels are $0$, $5$ and $10$, respectively. The results show that the degradation representation learning module can improve performance of our proposed BlindSRSNF on blind SR tasks.

\subsubsection{Sampling Interval}
\begin{table}[ht]
    \centering
    \renewcommand\arraystretch{1.4}
    \caption{Discussion on Sampling Interval of Reverse Process. ``\#'' Indicates the Setting Used in Our Experiments.}
    \label{tab:ablation-gamma}
    \setlength{\tabcolsep}{0.5mm}{
        \begin{tabular}{p{3em}|p{6em}<{\centering}p{6em}<{\centering}p{6em}<{\centering}p{6em}<{\centering}}
            \hline
            \hline
            $\gamma$ & PSNR(\si{\dB}) & FID   & LPIPS & Runtime (s) \bigstrut \\
            \hline
            1       & 20.25 & 70.71 & 0.1553 & 20.29  \bigstrut[t] \\
            10      & 20.40 & 70.74 & 0.1523 & 2.05 \\
            25      & 20.54 & 66.20 & 0.1514 & 0.83 \\
            50 (\#) & 20.82 & 66.65 & 0.1485 & 0.45 \\
            100     & 21.24 & 71.99 & 0.1661 & 0.25 \\
            200     & 21.71 & 91.19 & 0.2061 & 0.14 \\
            500     & 22.23 & 114.33 & 0.2832 & 0.09  \bigstrut[b] \\
            \hline
            \hline
        \end{tabular}%
    }
\end{table}

The sampling interval $\gamma$ of the reverse process is an important parameter of the SNF. It determines the number of times the denoising model is performed in the reverse process, so it will directly affect the inference time and the performance of our model. Table \ref{tab:ablation-gamma} shows the effect of sampling interval on model performance and runtime. Fig.\,\ref{fig:ablation-gamma} shows a comparison of visual results for different sampling intervals.

Table \ref{tab:ablation-gamma} shows that the PSNR keeps increasing with the sampling interval increasing. The proposed method achieves the best FID and LPIPS scores when the sampling interval is set to $25$ or $50$. When the sampling interval is greater than $100$, although a higher PSNR is obtained, the FID and LPIPS scores drop significantly. As can be seen from Fig.\,\ref{fig:ablation-gamma}, too small or too large sampling interval will lead to a decrease in the perception quality of SR results. From the results of $\gamma=1$, obvious pseudo textures appear in the farmland area; while the SR results of $\gamma=500$ are blurry and lack clear texture details. This is because a smaller sampling interval results in more sampling steps in reverse process of SNF, and the method will tend to generate more texture details. Although generating more textures can significantly improve the visual perceptual quality of SR results, the potential pseudo-textures can also exacerbate the spatial distortion of results, reflected as a decrease in PSNR values.

Furthermore, the sampling interval is inversely proportional to the runtime. Therefore, too small sampling interval will greatly increase the computational cost in the inference phase. In conclusion, to comprehensively balance the visual perception quality, runtime and spatial distortion degree of the BlindSRSNF, we set the sampling interval to $50$.

It is worth noting that the proposed method can control the richness of the texture by adjusting the sampling interval during the inference stage. This characteristic allows user to flexibly control the performance of the method during the inference phase without retraining the model, which is not possible with GAN-based models.

\section{Conclusions}
\label{sec:conclusion}

In this article, we propose a novel blind SR algorithm based on SNF to better handle various blur kernels and noise levels in real-world RSIs. The BlindSRSNF realizes the probability distribution transformation between the prior space and the target space through a Markov process. Combining the LR encoding and the degradation representation vector, we construct a conditional transition probability of the reverse diffusion process, which makes it possible to explicitly optimize the NLL of the generative model. This optimization mechanism significantly reduces the training difficulty of generative models compared to GAN-based algorithms. We propose to use pixel folding and pixel shuffle operations to reduce the dimension of feature maps, combined with the interval sampling strategy, which effectively improves the sampling efficiency of flow-based models. Furthermore, we introduce a contrastive learning-based degradation representation strategy to avoid the error amplification problem caused by inaccurate degradation kernel estimation. Comprehensive experiments on the GeoEye-1 and GoogleEarth datasets show that the BlindSRSNF improves the performance of blind SR compared to the SOTA algorithms. Visual results show that the proposed BlindSRSNF can more realistically restore the details of ground objects in real-world LR RSIs.
\balance


%

\appendices



\ifCLASSOPTIONcaptionsoff
    \newpage
\fi



\bibliographystyle{IEEEtran}
\bibliography{IEEEabrv,refs}

\end{document}